\documentclass[twocolumn,showpacs,preprintnumbers,amsmath,
               nofootinbib,aps,superscriptaddress]{revtex4}

\usepackage{graphicx}% Include figure files
\usepackage{dcolumn}% Align table columns on decimal point
\usepackage{bm}% bold math

\usepackage{ulem}% to strike out some text with \sout{...}
\usepackage[usenames]{color}% to use colored text

\newcommand*{\no}{\noindent}
\newcommand*{\bea}{\begin{eqnarray}}
\newcommand*{\eea}{\end{eqnarray}}
\newcommand*{\be}{\begin{equation}}
\newcommand*{\ee}{\end{equation}}
\newcommand*{\pd}{\partial}
\newcommand*{\pdm}{\pd_{\mu}}

\newcommand*{\pref}[1]{(\ref{#1})}

\newcommand*{\mn}{{\mu\nu}}

\newcommand*{\prefr}[2]{(\ref{#1}--\ref{#2})} 
\newcommand*{\nn}{\nonumber}

\newcommand*{\tr}{\mathrm{tr}}
\newcommand*{\diag}{\mathrm{diag}}

\newcommand*{\tl}{\mathrm{tl}}

\begin{document}

\preprint{}

\title{Yang-Mills Theory in $\lambda$-Gauges}

\author{Axel Maas}\email{axelmaas@web.de}
\affiliation{Theoretical-Physical Institute, Friedrich-Schiller-University Jena, Max-Wien-Platz 1, D-07743 Jena, Germany}

\author{Tereza Mendes}\email{mendes@ifsc.usp.br}
\affiliation{Instituto de F\'isica de S\~ao Carlos, University of S\~ao Paulo, \\
              Caixa Postal 369, 13560-970 S\~ao Carlos, SP, Brazil}

\author{\v{S}tefan Olejn{\'\i}k}\email{stefan.olejnik@savba.sk}
\affiliation{Institute of Physics, Slovak Academy of Sciences, SK--845 11 Bratislava, Slovakia}

\date{\today}

\begin{abstract}

The gauge-independent phenomenon of color confinement in Yang-Mills theory manifests 
itself differently in different gauges. 
Therefore, the gauge dependence of quantities related to the infrared 
structure of the theory becomes important for understanding the confinement mechanism. 
Particularly useful are classes of gauges that are controlled by a single gauge 
parameter. We present results on propagators and the color-Coulomb potential for 
the so-called $\lambda$-gauges, which interpolate between the (minimal) Landau gauge 
and the (minimal complete) Coulomb gauge. 
Results are reported for the SU(2) lattice gauge theory in three and four 
space-time dimensions. 
We investigate especially intermediate and low momenta. 
We find a continuous evolution of all quantities with the gauge parameter, 
except at zero four-momentum.

\end{abstract}

\pacs{11.15.Ha 12.38.Aw 14.70.Dj}
\maketitle

%%%%%%%%%%%%%%%%%%%%%%%%%%%%%%%%%%%%%%%%%%%%%%%%%%%%%%%%%%%%%%%%%%%%%%%%%%%%%%%%%%%%%%%%%%%%%%%%%%%

\section{Introduction}

The investigation of the confinement mechanism in QCD and Yang-Mills theory has progressed during the last decade,
but remains an unsolved and challenging problem 
(see e.g.\ \cite{Alkofer:2006fu} for a brief review).
A particularly important question is that of gauge dependence.

The confinement phenomenon has been investigated in various gauges, with different degrees of sophistication, methods, and observables. Still, at the present stage it is not yet clear to what extent the confinement mechanism itself is gauge-dependent. It is, however, sure that its manifestation in the structure of gauge-dependent quantities, like correlation functions, is different in different gauges.

For a full understanding of the confinement process, it is thus necessary to understand this mutability among different gauges. 
For such a task, classes of gauges are of special interest which offer parameters to deform the gauge condition smoothly. This allows one to track the changes of correlation functions along the gauge orbit.

One such class are the $\lambda$-gauges \cite{Baulieu:1998kx,Burnel:2008zz}, defined by the gauge condition
\bea
\pdm'A^{\mu a}(x)&=&0\label{gc}\,,\\
\pdm'&\equiv&(\lambda_0 \pd_0,\dots,\lambda_{d-1} \pd_{d-1})\nn\,,
\eea
\no where $d$ is the number of space-time dimensions and the $\lambda_\mu$'s 
are a set of gauge parameters. E.\ g., the Landau-gauge case will
be given by $\lambda_\mu=1$ for all $\mu$.
In principle, it is necessary to add a prescription for dealing with Gribov-Singer effects \cite{Gribov,Singer:dk}. We employ here the {\it minimal} extension of the 
perturbatively defined gauge \pref{gc} to the non-perturbative domain by selecting 
a random Gribov copy among all Gribov copies belonging to the first Gribov region. 
This region is defined by those gauge copies for which the associated Faddeev-Popov 
operator, defined below in equation \pref{fpop}, is non-negative.

Clearly, if one of the $\lambda_\mu$'s identically vanishes, the gauge condition 
is similar to a would-be Coulomb gauge, but has to be supplemented by further 
constraints to yield even a perturbatively complete gauge condition. 
In contrast, a complete Coulomb gauge condition is obtained 
by performing a limit $\lambda_\mu\to 0$ for a certain $\mu$ \cite{Cucchieri:2007uj}. 
It is not clear, whether the different lattice definitions of a perturbatively complete Coulomb gauge \cite{Cucchieri:2007uj,Burgio:2008jr} yield coinciding correlation functions. The results we present here indicate that this may be the case, at least for a subclass of these definitions. However, in case of sufficiently non-smooth completions of the Coulomb gauge some effects could be present \cite{Burgio:2008jr}.

Here, we will concentrate on Landau-like $\lambda$-gauges, in which we choose only one of the 
$\lambda_\mu$'s different from 1. As the lattice calculations presented here will 
be performed in Euclidean space-time at zero temperature, the direction is 
irrelevant, but will conventionally be chosen to be the time direction. 
Therefore we set $\lambda_0=\lambda$ and 
$\lambda_{1}=\lambda_2=\dots=\lambda_{d-1}=1$ henceforth.

In these gauges the simplest correlation functions are the gluon propagator and the 
Faddeev-Popov-ghost propagator. We will use here lattice gauge theory to 
determine these propagators in the case of SU(2) Yang-Mills theory. 
Furthermore, in Coulomb gauge there exists an additional color-Coulomb potential that bounds the conventional Wilson potential from above \cite{Zwanziger:2002sh}. 
A confining color-Coulomb potential is thus a necessary condition for confinement, and we shall also investigate how this potential develops from Landau to Coulomb gauge. 
There exists an associated order parameter, the so-called residual-gauge-symmetry order parameter \cite{Greensite:2004ke}, which signals whether the 
residual gauge symmetry that exists in Coulomb gauge is unbroken. 
Also its evolution with the gauge parameter $\lambda$ will be determined here.

A word of caution is due here. The behavior of the correlation functions in the asymptotic infrared domain has not yet been completely explained even in Landau gauge. 
In fact, despite great effort invested in analytic 
\cite{Braun:2007bx,Sorella:2009vt,Aguilar:2008xm,Zwanziger:2009je} and numerical \cite{Maas:2008ri,Cucchieri:2009zt,vonSmekal:2007ns,Sternbeck:2008mv} studies, a consistent description unifying the different approaches is still being elaborated. 
For recent reviews of analytic and lattice studies see respectively \cite{Fischer:2008uz,Dudal:2009bf,Maas:2011se} and \cite{Maas:2011se,Cucchieri:2010xr}. 
Since the gauge-fixing process in the interpolating gauges is even more computationally expensive than in ordinary Coulomb or Landau gauge, it will not be possible here to reach volumes permitting to answer the question whether a scaling-type or a decoupling-type behavior is observed for general $\lambda$-gauges. 
The main interest is therefore on the general deformation of the relevant quantities with the gauge parameter.
Note that functional-equation studies predict the existence of a scaling-type solution in these gauges in four dimensions \cite{Fischer:2005qe}. For three dimensions, this will be shown in the appendix. Following the lines of the argument given in \cite{Fischer:2008uz}, it is easily conceivable that also a decoupling type of solution exists in this class of gauges. 
In any case, we will not attempt to settle this question here.

The lattice definition of the $\lambda$-gauges and the details of the production of the corresponding configurations are outlined in Section \ref{sdef}. The results for the propagators are presented in Section \ref{sprop}. Definitions of the propagators are also introduced there. Results for three dimensions are given in Subsection \ref{s3d}, while those for four dimensions are given in Subsection \ref{s4d}. Definitions and results for the color-Coulomb potential and associated quantities will be discussed in Section \ref{scopot}. A summary is given in Section~\ref{ssum}.

The results presented in this paper extend earlier studies \cite{Cucchieri:2001tw}. In particular, we complement our previous investigation \cite{Cucchieri:2007uj}, in which a qualitative difference of some parts of the gluon propagator was observed in $d=4$. Some preliminary results were presented in Ref.\ \cite{Maas:2006fk}. See also \cite{Iritani:2011zg} for investigations of other setups and quantities in this type of gauges.

\section{Definitions and set-up}\label{sdef}

\subsection{Generation of gauge-fixed configurations}

Minimal $\lambda$-gauges on the lattice are a straightforward generalization of the conventional minimal Landau gauge \cite{Cucchieri:2001tw,Cucchieri:2003fb}. In particular, to fix the gauge, a gauge transformation $g(x)$ is determined, which minimizes the functional
\be
{\cal E}(g(x))=1-\frac{a^d}{2dV}\sum_{\mu}\lambda_\mu\sum_x
\tr\left[U_\mu^{g}(x)+U_\mu^{g+}(x)\right].\label{minfunc}
\ee
\no Here $a$ is the lattice spacing and $V=N^d$ the lattice volume. $U_\mu\in SU(N_c)$, with $N_c=2$ for the present SU(2) case, is a link variable in the lattice configuration of the gauge field, and its gauge transform $U_\mu^g$ is given by
\be
U_\mu^g(x)=g(x)U_\mu(x)g^+\left(x+a\vec{e}_\mu\right).\nn
\ee
\no For a more detailed discussion of this gauge condition, in particular for the limit of $\lambda\to 0$ and the connection to Coulomb gauge, see \cite{Cucchieri:2007uj}.

Minimizing the above functional guarantees that the gauge-fixed configuration is within the first Gribov horizon, defined by the first null eigenvalue of the lattice Faddeev-Popov operator
\bea
\! M^{ab}(x,y)\omega^b(y)\!\!&=&\!\!\delta_{xy}\sum_{\mu}\lambda_\mu\{G_\mu^{ab}(y)[\omega^b(y)-\omega^b(y+e_\mu)]\nn\\
&& \quad -G_\mu^{ab}(y-e_\mu)[\omega^b(y-e_\mu)-\omega^b(y)]\nn\\[2mm]
&& \quad +\sum_{c}f^{abc}[A_\mu^b(y)\omega^c(y+e_\mu) \nn\\
&& \quad \;\; -A_\mu^b(y-e_\mu)\omega^c(y-e_\mu)]\}\,,\label{fpop}\\
G^{ab}_\mu(x)&=&\frac{1}{8}\tr\left(\{\tau_a,\tau_b\}\left[U_\mu(x)+U_\mu(x)^+\right]\right),\nn\\
A_\mu^a&=&\frac{\sqrt{\beta}}{4ia}\tr\left( \tau^a (U_\mu-U_\mu^+)\right)+{\cal O}(a^2)\nn,
\eea
\no with $\tau^a$ the generators of SU(2). As said above, we select always a random copy from the set of Gribov copies satisfying the constraint in Eq.\ (\ref{gc}) through the minimization of the functional in Eq.\ (\ref{minfunc}), thereby implementing the so-called {\it minimal} version of all of these gauges. 

The functional \pref{minfunc} can be minimized using the same methods as in Landau gauge \cite{Cucchieri:2003fb}. In particular, a stochastic overrelaxation algorithm is used, analogous to the Landau-gauge case \cite{Cucchieri:2006tf}. One should note that the required number of gauge-fixing sweeps increases with decreasing $\lambda$. Hence simulations at smaller and smaller $\lambda$ become computationally more and more expensive. Although the number of gauge-fixing sweeps can be decreased by multiple updates of each individual time-slice before changing the time-slice, this does not decrease the total computing time needed, due to the increase of the number of updates on single time-slices.

Thus, we limit our systematic investigations here to $\lambda\ge 10^{-2}$. Nevertheless, checks on small volumes down to $\lambda=10^{-3}$ do not show any qualitatively new effects. 

We distinguish furthermore between three types of Coulomb gauge fixing:
\begin{itemize}
\item Simple Coulomb gauge, defined by setting $\lambda$ to 0.
\item Complete Coulomb gauge, defined by fixing the residual (perturbative) gauge freedom in the simple Coulomb gauge in a Landau-like manner \cite{Cucchieri:2007uj}. 
\item Limiting Coulomb gauge, defined by the limit $\lambda\to 0$. 
Note that while this limit is well-defined in the sense of a gauge condition, the individual Green's functions do not necessarily have a pointwise-smooth limit\footnote{However, our results suggest that the limit is smooth except at $p=0$.}. 
\end{itemize}
The last two versions of the Coulomb gauge should coincide in the limit of $\lambda\to 0$ \cite{Cucchieri:2007uj}. We will determine the gluon propagator and the color-Coulomb potential for these three different prescriptions of fixing the Coulomb gauge. In all cases, the results coincide, as far as resolvable. This will be discussed in more detail below.

\begin{table*}[ht]
\caption{\label{conf} Information on the configurations considered in our numerical simulations. The value ``0 fixed'' for $\lambda$ denotes the case when the residual gauge freedom (after setting $\lambda$ to 0) is fixed according to \cite{Cucchieri:2007uj}. The value of the lattice spacing $a$ has been taken from \cite{Cucchieri:2003di} for the three-dimensional case and from \cite{Fingberg:1992ju,Bloch:2003sk} for the four-dimensional case. {\it Sweeps} indicates the number of sweeps \cite{Cucchieri:2006tf} between two consecutive gauge-fixed measurements. In all cases 200 thermalization sweeps have been performed. More details on the generation of configurations, error-determination, etc.\ can be found in \cite{Cucchieri:2006tf}.}
\begin{ruledtabular}
\vspace{1mm}
\begin{tabular}{c c c c c c c c c c c c c}
&\multicolumn{6}{c}{$d=3$ dimensions}&\multicolumn{6}{c}{$d=4$ dimensions}\\
\cline{2-7}\cline{8-13}
$\lambda$  & $N$ & $\beta$ & $a^{-1}$ [GeV] & Conf. & Sweeps & $V^{1/d}$ [fm] & $N$ & $\beta$ & $a^{-1}$ [GeV] & Conf. & Sweeps & $V^{1/d}$ [fm]\cr
\hline
%\hline
1           & 40  & 4.2     & 1.136          & 586   & 50     & 6.9               & 22  & 2.2     & 0.938          & 425   & 50     & 4.6           \cr
%\hline
1           & 40  & 6.0     & 1.733          & 689   & 50     & 4.5               & 22  & 2.5     & 2.309          & 532   & 50     & 1.9           \cr
%\hline
1         &     &     &         &                &       &                        & 22  & 2.8     & 5.930          & 342   & 50     & 0.73          \cr
%\hline
1         &     &     &         &                &       &                        & 40  & 2.2     & 0.938          & 50    &        & 8.4           \cr
\hline
%\hline
1/2         & 40  & 4.2     & 1.136          & 466   & 50     & 6.9               & 22  & 2.2     & 0.938          & 380   & 50     & 4.6           \cr
%\hline
1/2         & 40  & 6.0     & 1.733          & 542   & 50     & 4.5               & 22  & 2.5     & 2.309          & 372   & 50     & 1.9           \cr
%\hline
1/2       &     &     &         &                &            &                   & 22  & 2.8     & 5.930          & 362   & 50     & 0.73          \cr
\hline
%\hline
1/10        & 20  & 4.2     & 1.136          & 481   & 30     & 3.5             &     &         &                &       &        &               \cr
%\hline
1/10        & 20  & 6.0     & 1.733          & 555   & 30     & 2.3             &     &         &                &       &        &               \cr
%\hline
1/10        & 40  & 4.2     & 1.136          & 424   & 50     & 6.9               & 22  & 2.2     & 0.938          & 311   & 50     & 4.6           \cr
%\hline
1/10        & 40  & 6.0     & 1.733          & 461   & 50     & 4.5               & 22  & 2.5     & 2.309          & 370   & 50     & 1.9           \cr
%\hline
1/10      &     &     &         &                &       &                        & 22  & 2.8     & 5.930          & 399   & 50     & 0.73          \cr
%\hline
1/10        & 60  & 4.2     & 1.136          & 314   & 70     & 10                & 40  & 2.2     & 0.938          & 50    &        & 8.4           \cr
%\hline
1/10        & 60  & 6.0     & 1.733          & 322   & 70     & 6.8               & 70  & 2.2     & 0.938          & 310   &        & 14.7          \cr
\hline
%\hline
1/20        & 40  & 4.2     & 1.136          & 447   & 50     & 6.9               & 22  & 2.2     & 0.938          & 364   & 50     & 4.6           \cr
%\hline
1/20        & 40  & 6.0     & 1.733          & 463   & 50     & 4.5               & 22  & 2.5     & 2.309          & 307   & 50     & 1.9           \cr
%\hline     
1/20      &     &     &         &                &       &                        & 22  & 2.8     & 5.930          & 409   & 50     & 0.73          \cr
%\hline
1/20      &     &     &         &                &       &                        & 40  & 2.2     & 0.938          & 50    &        & 8.4           \cr
%\hline
1/20      &     &     &         &                &       &                        & 70  & 2.2     & 0.938          & 336   &        & 14.7          \cr
\hline
%\hline
1/100     &     &     &         &                &       &                        & 14  & 2.2     & 0.938          & 418   & 30     & 2.9           \cr
%\hline
1/100     &     &     &         &                &       &                        & 14  & 2.5     & 2.309          & 420   & 30     & 1.2           \cr
%\hline
1/100       & 40  & 4.2     & 1.136          & 356   & 50     & 6.9               & 22  & 2.2     & 0.938          & 262   & 50     & 4.6           \cr
%\hline
1/100       & 40  & 6.0     & 1.733          & 478   & 50     & 4.5               & 22  & 2.5     & 2.309          & 367   & 50     & 1.9           \cr
%\hline
1/100     &     &     &         &                &       &                        & 22  & 2.8     & 5.930          & 332   & 50     & 0.73          \cr
%\hline
1/100     &     &     &         &                &       &                        & 32  & 2.2     & 0.938          & 50    &        & 6.7           \cr
%\hline
1/100     &     &     &         &                &       &                        & 40  & 2.2     & 0.938          & 50    &        & 8.4           \cr
%\hline
1/100     &     &     &         &                &       &                        & 70  & 2.2     & 0.938          & 116   &        & 14.7          \cr
\hline
%\hline
0           & 40  & 4.2     & 1.136          & 375   & 50     & 6.9               & 22  & 2.2     & 0.938          & 327   & 50     & 4.6           \cr
%\hline
0           & 40  & 6.0     & 1.733          & 352   & 50     & 4.5               & 22  & 2.5     & 2.309          & 340   & 50     & 1.9           \cr
%\hline
0         &     &     &         &                &       &                        & 22  & 2.8     & 5.930          & 319   & 50     & 0.73          \cr
\hline
%\hline
0 fixed     & 40  & 4.2     & 1.136          & 377   & 50     & 6.9               & 22  & 2.2     & 0.938          & 434   & 50     & 4.6           \cr 
%\hline
0 fixed     & 40  & 6.0     & 1.733          & 314   & 50     & 4.5               & 22  & 2.5     & 2.309          & 309   & 50     & 1.9           \cr
%\hline
0 fixed   &     &     &         &                &       &                        & 22  & 2.8     & 5.930          & 401   & 50     & 0.73          \cr
%\hline
\end{tabular}
\end{ruledtabular}
\end{table*}

\begin{table}[ht]
\caption{\label{conf2}Configurations used for the determination of the color-Coulomb potential and the residual-gauge-symmetry order parameter. For details, see Table \ref{conf}. All calculations have been performed in four dimensions. The columns with various $\beta$'s indicate the number of configurations used. The number of sweeps was always 50.}
\begin{ruledtabular}
\vspace{1mm}
\begin{tabular}{c c c c c c}
%\hline
          &     &             &             &             & $V^{1/d}$ [fm]\cr
$\lambda$ & $N$ & $\beta=2.2$ & $\beta=2.5$ & $\beta=2.8$ & at $\beta=2.2$ \cr
\hline
1         & 4   & 188         & 188         & 188         & 0.84                          \cr
%\hline
1         & 6   & 190         & 190         & 190         & 1.3                           \cr
%\hline
1         & 8   & 191         & 191         & 191         & 1.7                           \cr
%\hline
1         & 10  & 192         & 192         & 192         & 2.1                           \cr
%\hline
1         & 12  & 193         & 193         & 193         & 2.5                           \cr
%\hline
1         & 16  & 195         & 195         & 195         & 3.4                           \cr
%\hline
1         & 20  & 196         & 196         & 196         & 4.2                           \cr
%\hline
1         & 24  & 104         &             & 197         & 5.0                           \cr
\hline
1/10      & 4   & 188         & 188         & 188         & 0.84                          \cr
%\hline
1/10      & 6   & 190         & 190         & 190         & 1.3                           \cr
%\hline
1/10      & 8   & 191         & 191         & 191         & 1.7                           \cr
%\hline
1/10      & 10  & 192         & 192         & 192         & 2.1                           \cr
%\hline
1/10      & 12  & 193         & 193         & 193         & 2.5                           \cr
%\hline
1/10      & 16  & 195         & 195         & 195         & 3.4                           \cr
%\hline
1/10      & 20  & 180         & 196         & 196         & 4.2                           \cr
%\hline
1/10      & 24  & 94          & 298         & 197         & 5.0                           \cr
\hline
1/100     & 4   & 188         & 188         & 188         & 0.84                          \cr
%\hline
1/100     & 6   & 190         & 190         & 190         & 1.3                           \cr
%\hline
1/100     & 8   & 191         & 191         & 191         & 1.7                           \cr
%\hline
1/100     & 10  & 192         & 192         & 192         & 2.1                           \cr
%\hline
1/100     & 12  & 193         & 193         & 193         & 2.5                           \cr
%\hline
1/100     & 16  & 100         & 195         & 195         & 3.4                           \cr
%\hline
1/100     & 20  & 40          & 192         & 138         & 4.2                           \cr
%\hline
1/100     & 24  & 29          & 197         & 197         & 5.0                           \cr
\hline
0         & 4   & 188         & 188         & 188         & 0.84                          \cr
%\hline
0         & 6   & 190         & 190         & 190         & 1.3                           \cr
%\hline
0         & 8   & 191         & 191         & 191         & 1.7                           \cr
%\hline
0         & 10  & 192         & 192         & 192         & 2.1                           \cr
%\hline
0         & 12  & 193         & 193         & 193         & 2.5                           \cr
%\hline
0         & 16  & 195         & 195         & 195         & 3.4                           \cr
%\hline
0         & 20  & 115         & 196         & 196         & 4.2                           \cr
%\hline
0         & 24  & 70          & 197         & 197         & 5.0                           \cr
\hline
0 fixed   & 4   & 188         & 188         & 188         & 0.84                          \cr
%\hline
0 fixed   & 6   & 190         & 190         & 190         & 1.3                           \cr
%\hline
0 fixed   & 8   & 191         & 191         & 191         & 1.7                           \cr
%\hline
0 fixed   & 10  & 192         & 192         & 192         & 2.1                           \cr
%\hline
0 fixed   & 12  & 193         & 193         & 193         & 2.5                           \cr
%\hline
0 fixed   & 16  & 195         & 195         & 195         & 3.4                           \cr
%\hline
0 fixed   & 20  & 82          & 196         & 196         & 4.2                           \cr
%\hline
0 fixed   & 24  & 90          & 197         & 197         & 5.0                           \cr
%\hline
\end{tabular}
\end{ruledtabular}
\end{table}

The details of the configurations employed for the determination of the propagator are given in Table \ref{conf}. Not all configurations have been used in the figures. The remaining ones have been obtained to investigate various aspects, such as finite-size effects. In these cases, the results are just stated in the text. The configurations used for the calculation of the color-Coulomb potential and associated quantities are listed in Table \ref{conf2}.

\section{Propagators in $\lambda$-gauges}\label{sprop}

Setting $\lambda$ to a value different from 1 yields a gauge condition \pref{gc} that is no longer manifestly Lorentz (Euclidean) invariant. Thus, as in Coulomb gauge, the energy $p_0$ and the spatial momentum $|\vec p|$ have to be considered as independent variables.

Furthermore, the gluon is a vector particle. Hence, it is no longer possible to describe it by a single scalar function as in Landau gauge, but two independent ones are required. For practical purposes we choose these two independent functions to be\footnote{Note that these are essentially the same definitions as were used in \cite{Cucchieri:2007uj}, since there only the case $p_0=0$ has been investigated. This definition has been chosen for numerical convenience. See appendix \ref{adse} for the relation to the definition of \cite{Fischer:2005qe} and a general decomposition.}
\bea
D_{00}(p_0,|\vec p|)&=&\frac{1}{(N_c^2-1)}<A_0^a A_0^a>\,,\label{d00}\\
D^\tr(p_0,|\vec p|)&=&\frac{1}{(N_c^2-1)(d-2)}\nn\\
&&\times\left(\delta_{ij}-\frac{p_i p_j}{p_0^2+\vec p^2}\right)<A_i^a A_j^a>\label{dtr},
\eea
\no where $i$ and $j$ run only over spatial indices. $D^\tr$ is evaluated at zero $d-1$-momentum, and thus also at zero $d$-momentum, by setting the second term in the above projector to zero and multiplying by $(d-2)/(d-1)$. This yields a unique and smooth limit from any direction for vanishing $d$-momentum. Note that at $\lambda=0$, the second term in \pref{dtr} is not contributing, since in this case $p_i A_i=0$, due to the Coulomb gauge condition. The relation to the conventional choice in DSE studies \cite{Fischer:2005qe} is detailed in Appendix \ref{adse}.

Herein, the gluon field $A_\mu^a$ in momentum space is obtained from the position-space variables by
\be
A_\mu^a(p)=e^{- \frac{i\pi P_{\mu}}{N}} \sum_X e^{2\pi i \sum_\mu \frac{P_\mu X_\mu}{N}} A_\mu^a(x) ,\nn
\ee
\no where on a finite lattice the components $P_\mu$ of $P$ have the integer values $-N/2 + 1\, , \ldots , \, N/2 \, $ and the components $X_\mu$ of $X$ are the (integer) coordinates in the lattice ranging from $0$ to $N-1$. The physical momenta $p$ to which the integer lattice ones $P$ correspond are
\be
p_\mu=\frac{2}{a}\sin\frac{P_\mu\pi}{N}\nn.
\ee

The ghost propagator $D_G(p_0,|\vec p|)$ is, as in Landau gauge, defined by the Fourier transform of the inverse Faddeev-Popov operator \pref{fpop}. As this operator is symmetric, the inversion can be done using a conjugate gradient method, as in ordinary Landau gauge \cite{Cucchieri:2006tf}. To determine the renormalization effects in four dimensions, it is furthermore useful to define the dressing function $G$ via
\be
D_G(p_0,|\vec p|)=\frac{1}{\lambda p_0^2+\vec p^2}G(p_0,|\vec p|).\label{ghp}
\ee
\no Thereby the tree-level part $1/(\lambda p_0^2+\vec p^2)$ of the propagator is explicitly factored out.

\subsection{Propagators in 3 dimensions}\label{s3d}

In three space-time dimensions, Yang-Mills theory in $\lambda$-gauges is, as in Landau gauge, finite on the level of correlation functions. Hence no renormalization effects occur. In particular, the gauge-parameter $\lambda$ is not different from its tree-level value.

It is worthwhile to study the three different propagators, the one of the transverse gluon \pref{dtr}, the one of the temporal gluon \pref{d00}, and the one of the ghost \pref{ghp}, individually.

\begin{figure*}
\includegraphics[width=\linewidth]{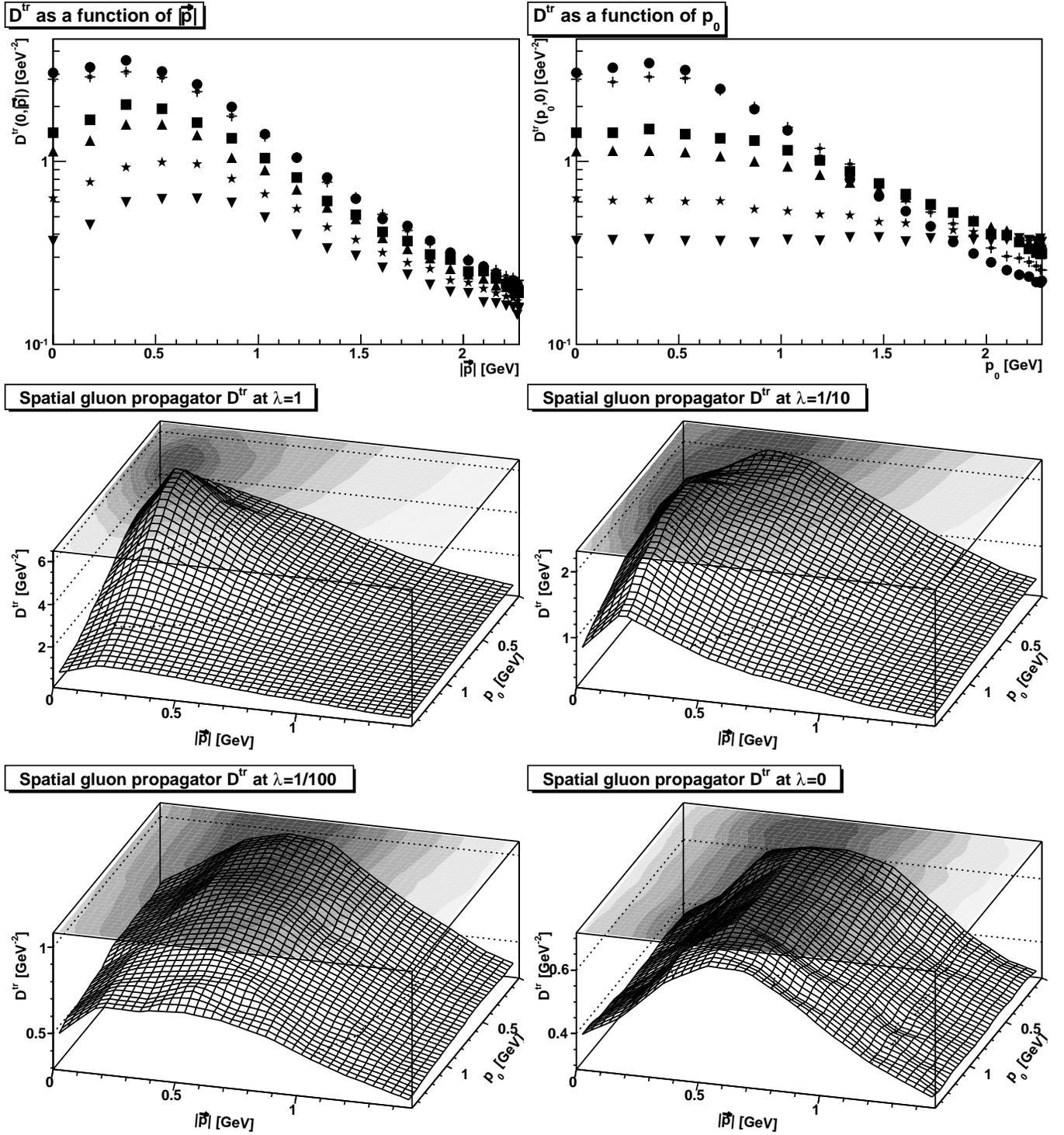}
\caption{\label{fdtr} The transverse gluon propagator $D^\tr$ for different values of $\lambda$ and different kinematic configurations, in the $3d$ case on a $40^3$ lattice. The top-left panel is the propagator as a function of pure spatial momenta $\vec p$, the top-right panel of pure temporal momenta $p_0$. The spatial momenta are measured along the $x$-axis. Results shown are for $\lambda=1$, i.e.\ Landau gauge (circles), $\lambda=1/2$ (crosses), $\lambda=1/10$ (squares), $\lambda=1/20$ (triangles), $\lambda=1/100$ (stars) and $\lambda=0$ (upside-down triangles). The lower panels show the full momentum dependence at $\lambda=1$ (middle-left panel), $\lambda=1/10$ (middle-right panel), $\lambda=1/100$ (bottom-left panel), and $\lambda=0$ (bottom-right panel). All results are at $\beta=4.2$. Also, $\lambda=0$ corresponds to the simple-Coulomb-gauge case. Results at $\beta=6.0$ look similar, but are more strongly affected by finite-size effects. (Note that scales on vertical axes are different in individual sub-figures.)}
\end{figure*}

The {\bf transverse gluon propagator} is shown for various kinematic configurations and values of $\lambda$ in Figure~\ref{fdtr}. There are a number of interesting observations, outlined next.

\begin{figure}
\includegraphics[width=\linewidth]{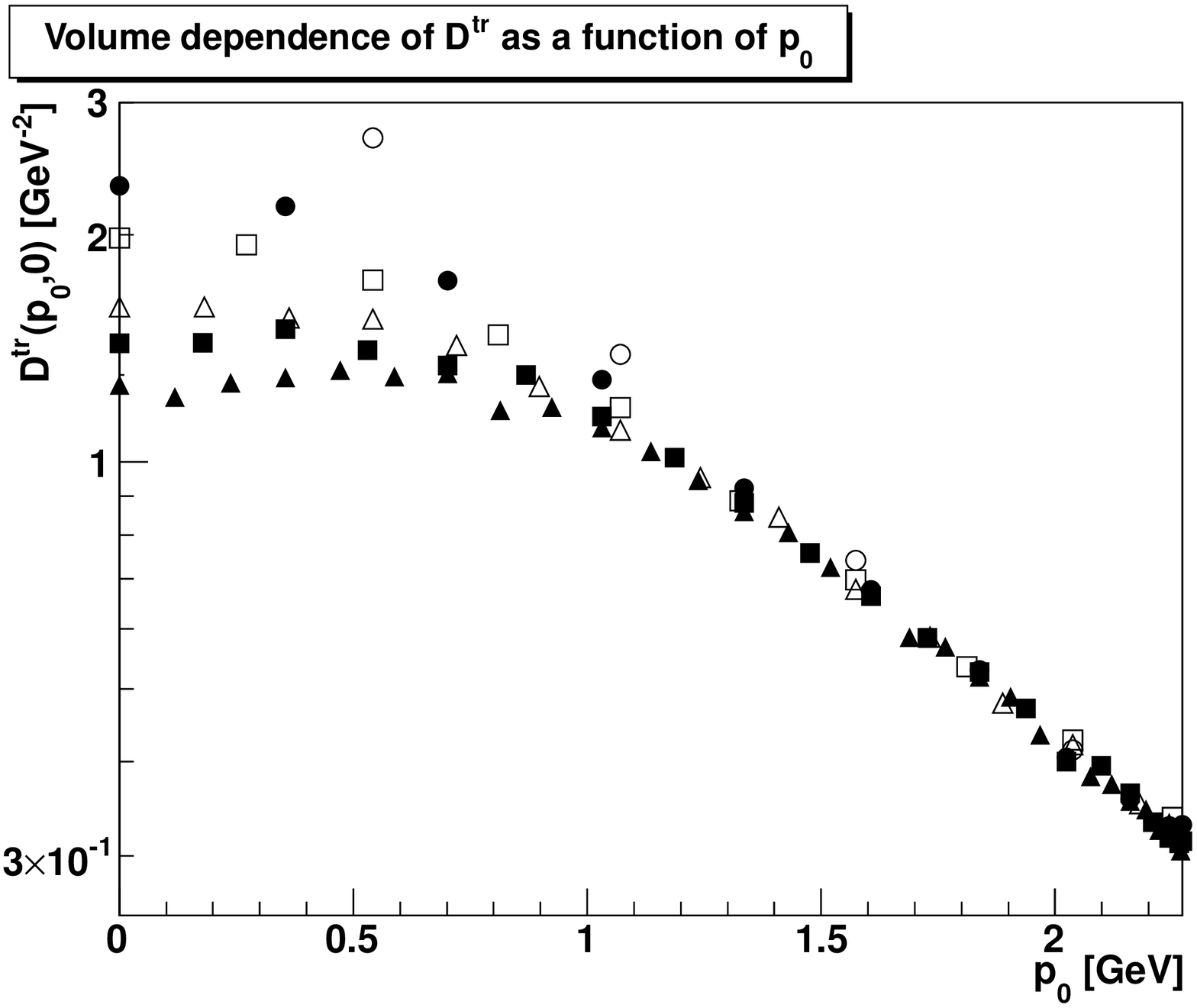}
\caption{\label{fdtr-v} The transverse gluon propagator $D^\tr$ as a function of $p_0$ at $\lambda=1/10$
for different volumes, in the $3d$ case. Open and full symbols correspond to $\beta=6.0$ and $\beta=4.2$, respectively. The
lattice sizes are $20^3$ (circles), $40^3$ (squares), and $60^3$ (triangles).}
\end{figure}

With respect to the spatial momentum, the clearly distinct maximum of the transverse gluon propagator in Landau gauge moves towards larger momenta with decreasing $\lambda$, albeit rather slowly. As for the dependence on the temporal momentum, the dependence on $p_0$ becomes flat at low momenta and the flat region becomes more extended when decreasing $\lambda$. However, the $p_0=0$ point is always lower, and may be influenced differently by finite-size effects. Already at $\lambda=1/10$, the maximum seems to be gone. The propagator at $\lambda=1/100$ is even more similar to the one of a massive particle. This is, however, a finite-volume effect. A comparison of different volumes, presented in Figure \ref{fdtr-v}, shows that the volumes in Figure \ref{fdtr} are too small to resolve the position of the maximum at $\lambda\lesssim 1/10$. It is visible that the maximum reappears at smaller momenta on a larger volume. Corresponding calculations to track the maximum at even lower values of $\lambda$ become, however, exceedingly costly.

For all three types of Coulomb gauge the transverse gluon propagators as functions of spatial momentum coincide. This is not surprising. Quantities that are only defined on a given time-slice, such as the spatial gluon propagator, should be independent of how the inter-time-slice gauge degree of freedom is fixed. Therefore, in the figures only one case, that of the simple Coulomb gauge, is shown.

We thus see that, in $3d$ Coulomb gauge, the transverse propagator for spatial momenta is similar to the one obtained in four dimensions in simple Coulomb gauge \cite{Cucchieri:2000gu}. It has a maximum and is infrared-suppressed.

On the other hand, the transverse propagator as a function of energy in Coulomb gauge is approximately constant. In the simple Coulomb gauge this is a consequence of the unfixed residual gauge degree of freedom. That this persists even in the case of fully fixed Coulomb gauges is not trivial \cite{Cucchieri:2007uj}, but has been deduced also in functional studies \cite{Reinhardt:2008pr}. In any case, in the alternative ways to fix the remaining gauge freedom, the gauge condition cannot impose a strict constraint on the gluon fields on any given configuration\footnote{Consider, e.g., $\pd_0 A_0=0$. Such a constraint is only fulfilled on average, $<\pd_0 A_0>=0$. The integrated constraint $\pd_0 \int d^{d-1}x_i A_0(x_0,x_i)=0$ is, however, true for any individual configuration.}. Hence, the fields still wash out when averaged over configurations, leaving only an instantaneous propagator. For this reason, the propagator is constant as a function of the energy. This is observed in all three types of Coulomb gauge and shown in Figure~\ref{fdtr} for the simple-Coulomb case. Note that this argument does not apply to the zero-energy mode. This mode coincides with the zero mode depending on spatial momenta, leading to the same value at zero $d$-momentum for the propagator, irrespective of the dependence on spatial momentum or on energy. If the propagator as a function of spatial momentum thus vanishes in the infinite-volume limit, it may have a discontinuity at zero as a function of energy.

Concluding, the spatial propagator $D^\tr(p)$ is instantaneous and infrared-suppressed in all versions of the Coulomb gauge considered here.

Although not shown, a further observation is that for the spatial propagator as a function of $\vec p$ the dependence on $\beta$ at fixed lattice size weakens with decreasing $\lambda$. However, at the same time the $\beta$-dependence increases for momenta along the time direction. This can indicate either a change in the volume dependence of the propagators with a change in $\lambda$, or a different $\beta$-dependence. The former is more likely, as can be seen in Figure \ref{fdtr-v}: The volume dependence is stronger than in corresponding calculations in Landau gauge in three dimensions \cite{Cucchieri:2003di,Cucchieri:2006tf}. It is furthermore visible that the $\beta$-dependence diminishes with increasing volume, making it more probable that this is a finite-volume effect, rather than a finite-$\beta$ effect.

\begin{figure*}
\begin{minipage}[c]{0.5\linewidth}
\includegraphics[width=0.96\linewidth]{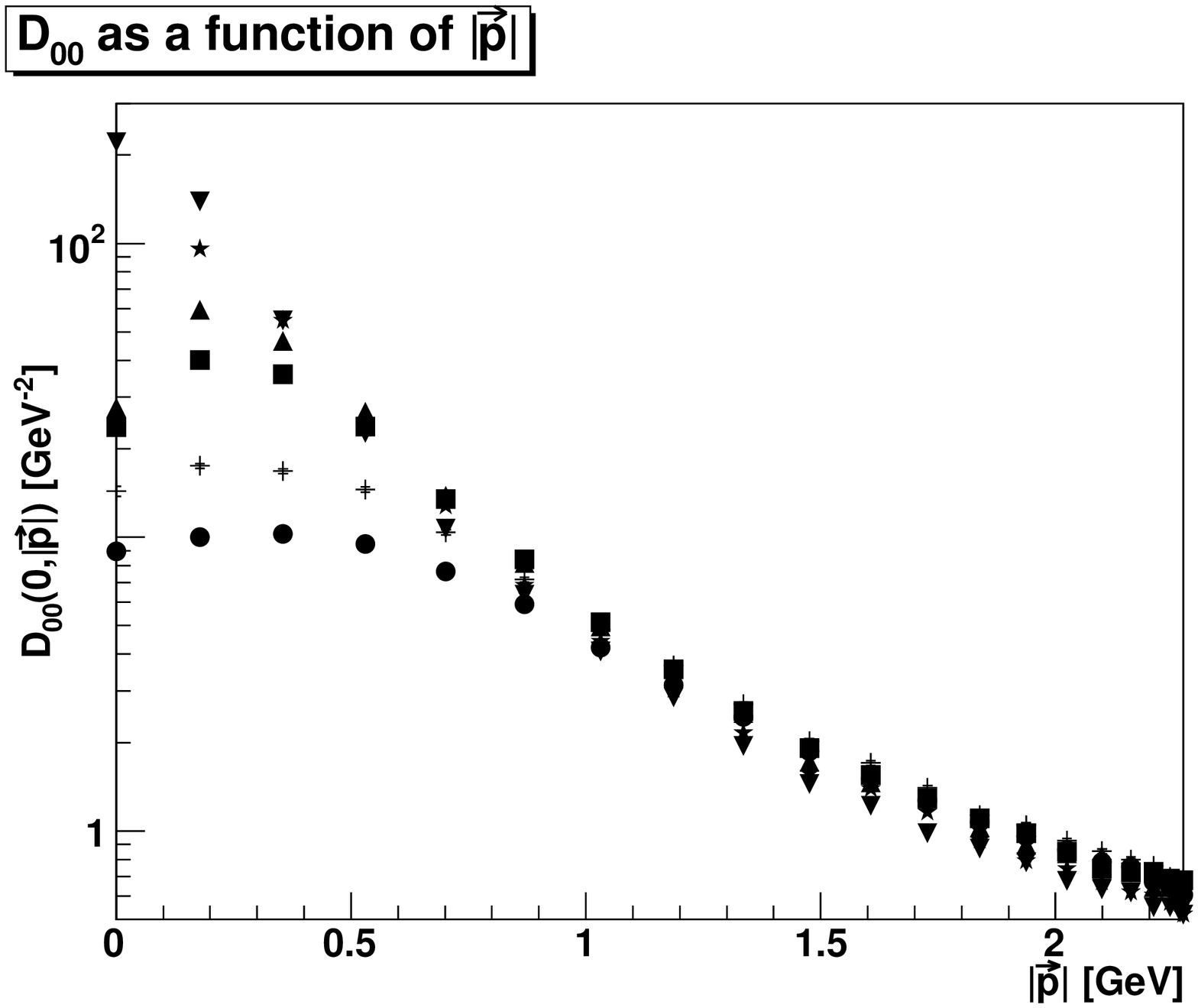}
\end{minipage}%
\begin{minipage}[c]{0.5\linewidth}
\caption{\label{fd00}The temporal gluon propagator $D_{00}$ for different values of $\lambda$ and different kinematic configurations, in the $3d$ case on a $40^3$ lattice. The top-left panel is the propagator as a function of pure spatial momenta $\vec p$. The momenta are measured along the $x$-axis. Results shown are for $\lambda=1$, i.e.\ Landau gauge (circles), $\lambda=1/2$ (crosses), $\lambda=1/10$ (squares), $\lambda=1/20$ (triangles), $\lambda=1/100$ (stars), and $\lambda=0$ (upside-down triangles). The panels below show the full momentum dependence at $\lambda=1$ (middle-left panel), $\lambda=1/10$ (middle-right panel), $\lambda=1/100$ (bottom-left panel), and $\lambda=0$ (bottom-right panel). All results are at $\beta=4.2$. Also, $\lambda=0$ corresponds to the simple-Coulomb-gauge case. Results at $\beta=6.0$ look similar, but are more strongly affected by finite-size effects. (Note that scales on vertical axes are different in individual sub-figures.)}
\end{minipage}\\
\begin{minipage}[c]{\linewidth}
\includegraphics[width=\linewidth]{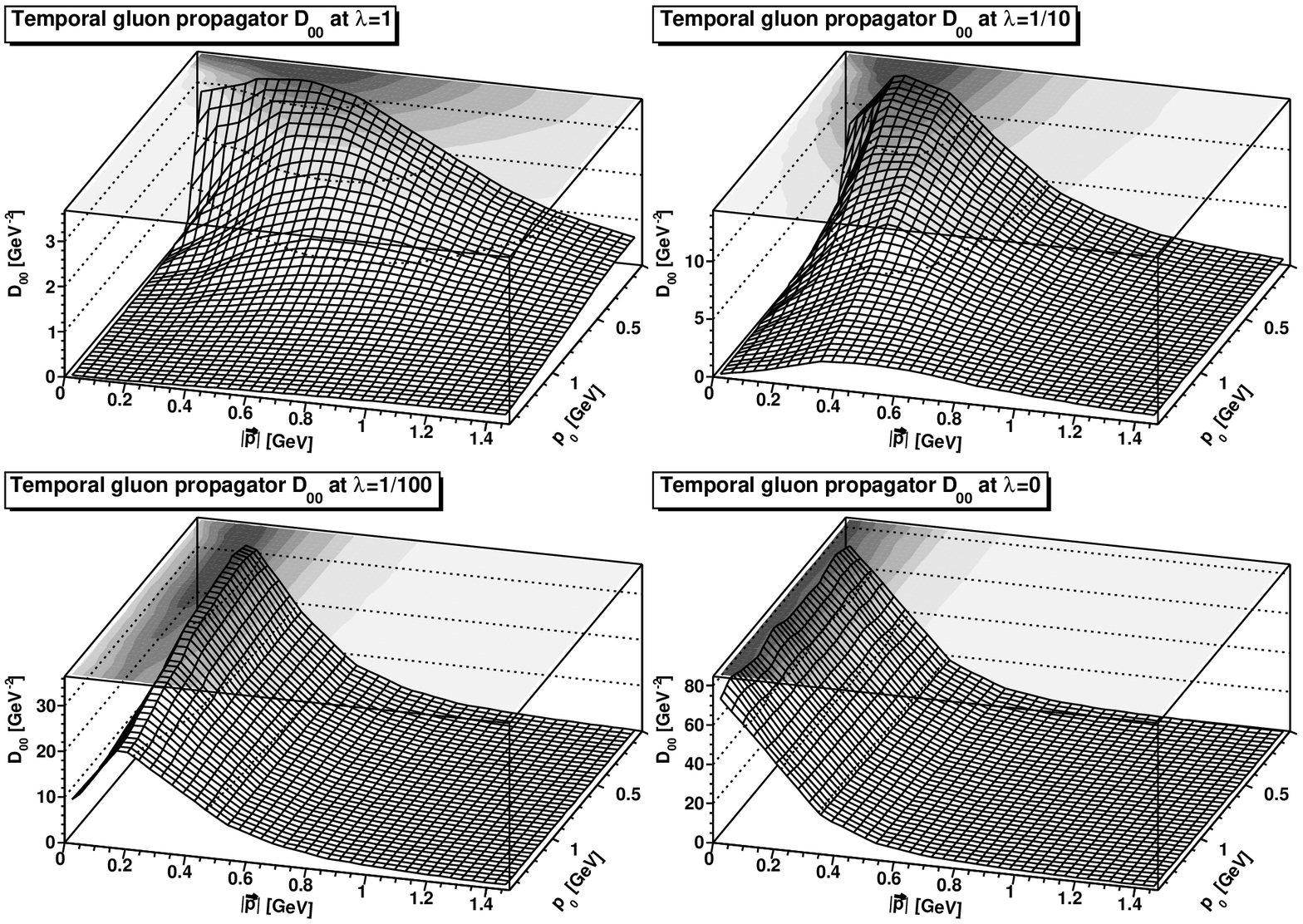}
\end{minipage}
\end{figure*}

\begin{figure}
\includegraphics[width=\linewidth]{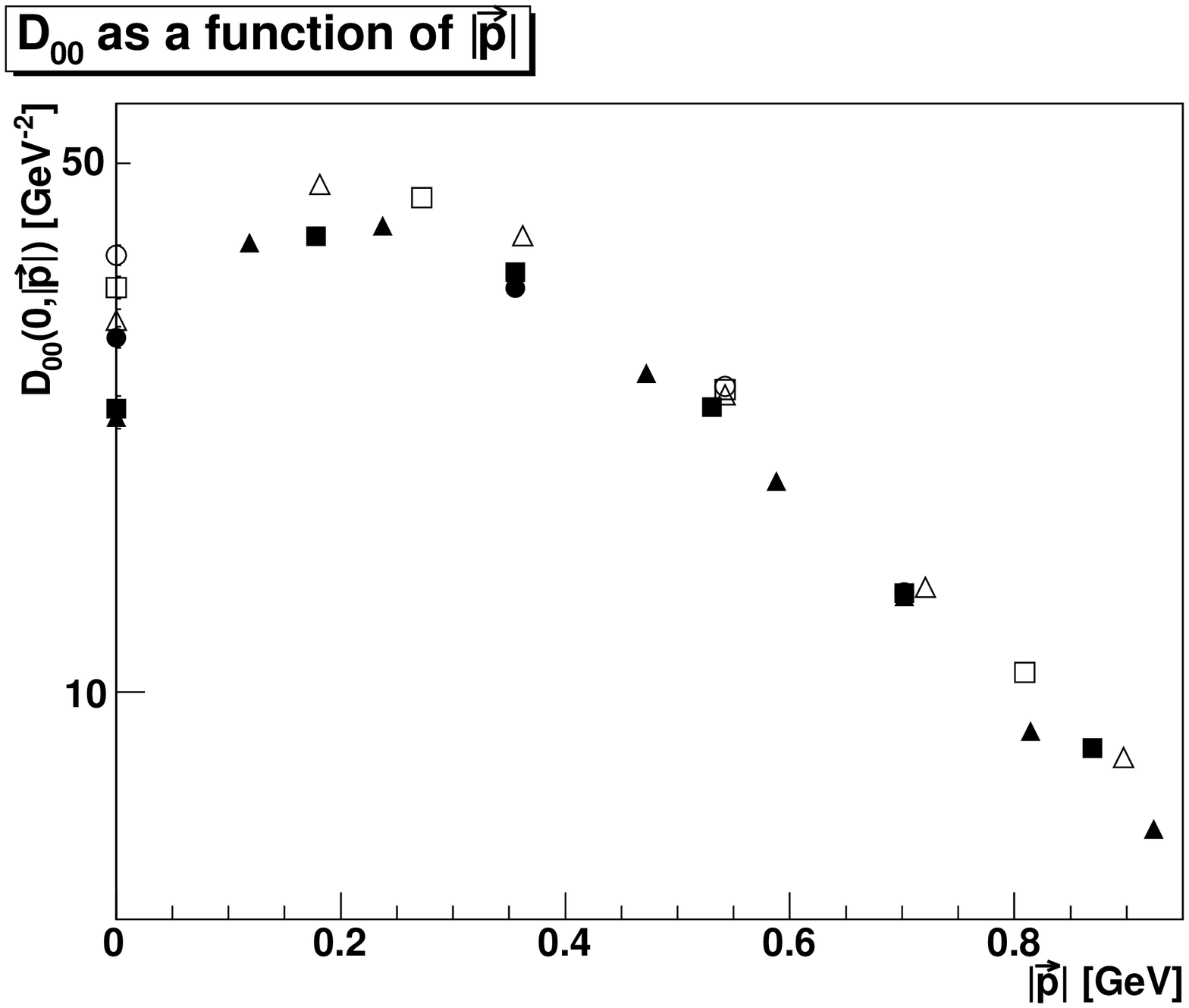}
\caption{\label{fd00-v} The temporal gluon propagator $D_{00}$ as a function of $|\vec p|$ at $\lambda=1/10$ for different volumes, in the $3d$ case. Open and full symbols correspond to $\beta=6.0$ and $\beta=4.2$, respectively. The lattice sizes are $20^3$ (circles), $40^3$ (squares), and $60^3$ (triangles).}
\end{figure}

The case of the {\bf temporal propagator} is at the first glance a bit simpler, as $D_{00}(p_0,0)$ vanishes identically for $p_0\neq 0$ due to the gauge condition. The remaining results are shown in Figure \ref{fd00}. As a function of spatial momenta, the maximum moves towards smaller momenta with decreasing $\lambda$. Again, already at $\lambda=1/10$, resolving the maximum is no longer possible for these volumes. At the smallest values of $\lambda$, the propagator looks more like a divergent function, rather than an infrared-suppressed function, except at the zero-momentum point for finite $\lambda$. However, this is again only a finite-volume effect, as is visible in Figure \ref{fd00-v}. This changes drastically for the Coulomb gauges, where the propagator diverges for $\vec p\to\vec 0$, independently of $p_0$, as it does already in perturbation theory. Thus, for all three definitions of Coulomb gauge the infrared-divergent gluon propagator is recovered. Note in particular that also the value at $p=0$ is non-vanishing in the Coulomb limit. Furthermore, the limit $\lambda\to 0$ is thus discontinuous for vanishing spatial momentum. It is, however, smooth for any non-vanishing spatial momentum. Concluding, the propagator is found to be instantaneous in all three Coulomb-gauge definitions.

\begin{figure*}
\begin{minipage}[c]{\linewidth}
\includegraphics[width=0.96\linewidth]{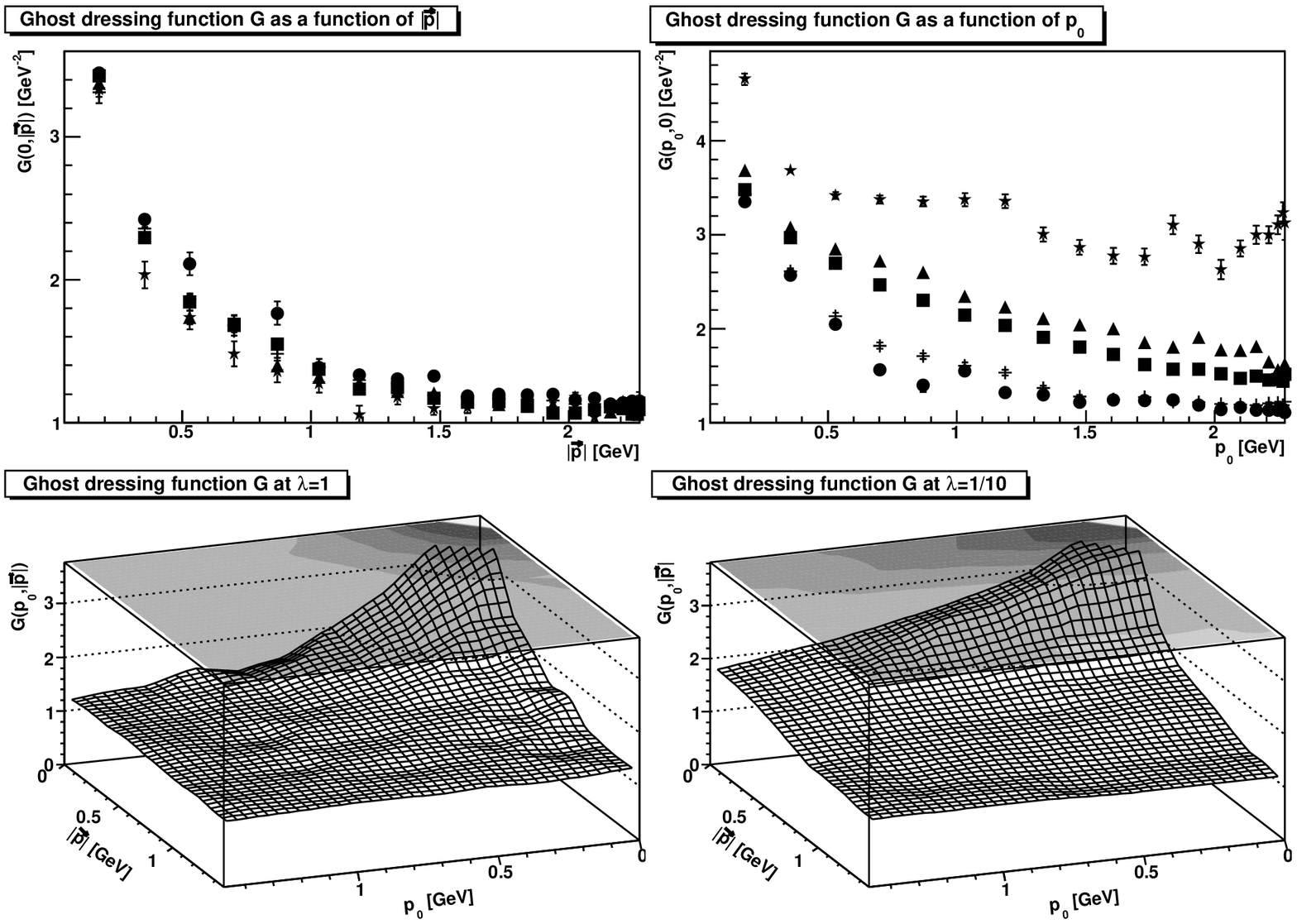}
\end{minipage}\\
\begin{minipage}[c]{0.5\linewidth}
\includegraphics[width=0.96\linewidth]{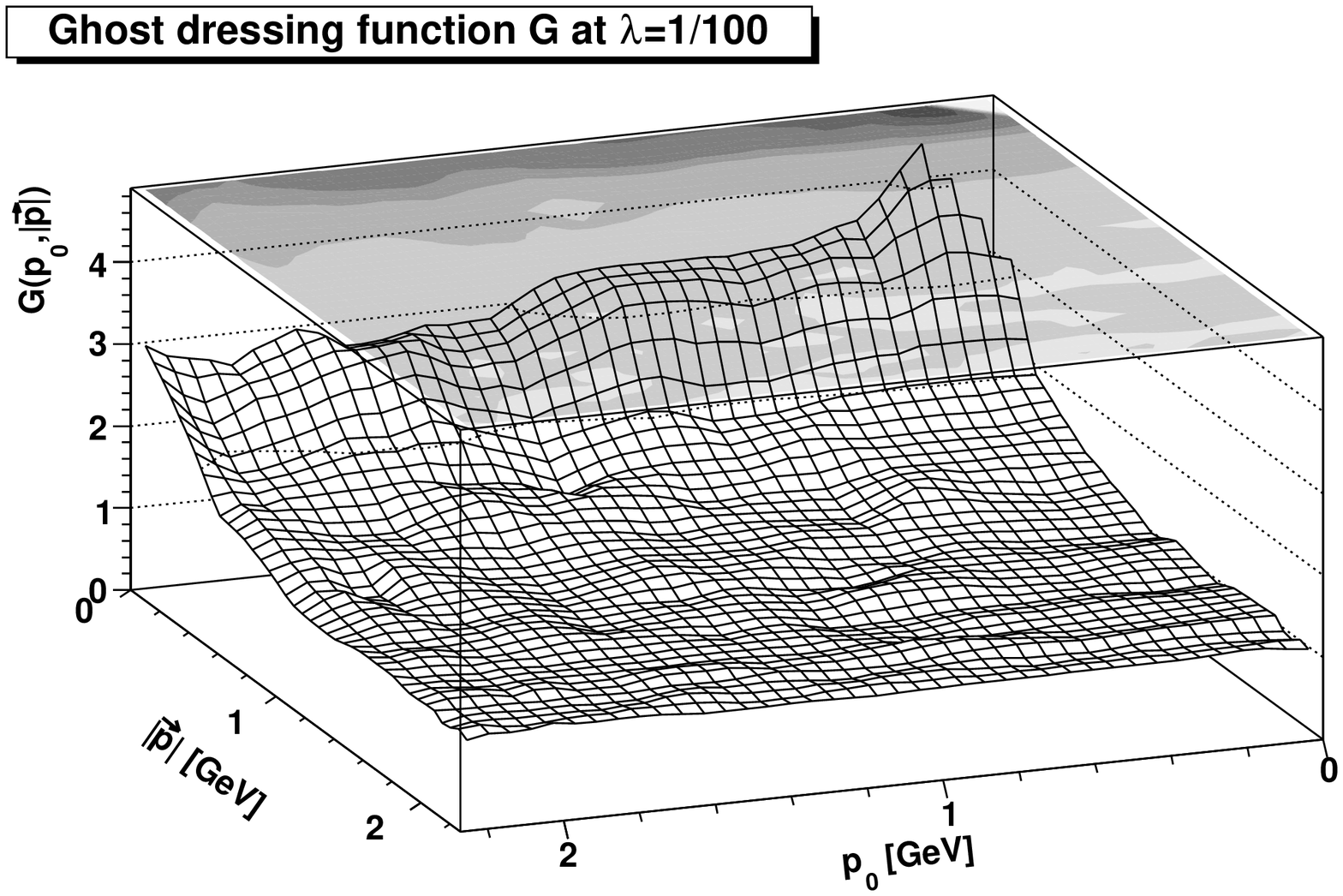}
\end{minipage}%
\begin{minipage}[c]{0.5\linewidth}
\caption{\label{fghp} The ghost dressing function $G$ for different values of $\lambda$ and different kinematic configurations, in the $3d$ case on a $40^3$ lattice. The top-left panel is the propagator as a function of pure spatial momenta $\vec p$, the top-right panel of pure temporal momenta $p_0$. The spatial momenta are measured along the $x$-axis. Results shown are for $\lambda=1$, i.e.\ Landau gauge (circles), $\lambda=1/2$ (crosses), $\lambda=1/10$ (squares), $\lambda=1/20$ (triangles), and $\lambda=1/100$ (stars). The lower panels show the full momentum dependence at $\lambda=1$ (middle-left panel), $\lambda=1/10$ (middle-right panel), and $\lambda=1/100$ (bottom-left panel). All results are at $\beta=4.2$. Also, $\lambda=0$ corresponds to the simple-Coulomb-gauge case.
Results at $\beta=6.0$ are essentially identical. }
\end{minipage}\\
\end{figure*}

\begin{figure}
\includegraphics[width=\linewidth]{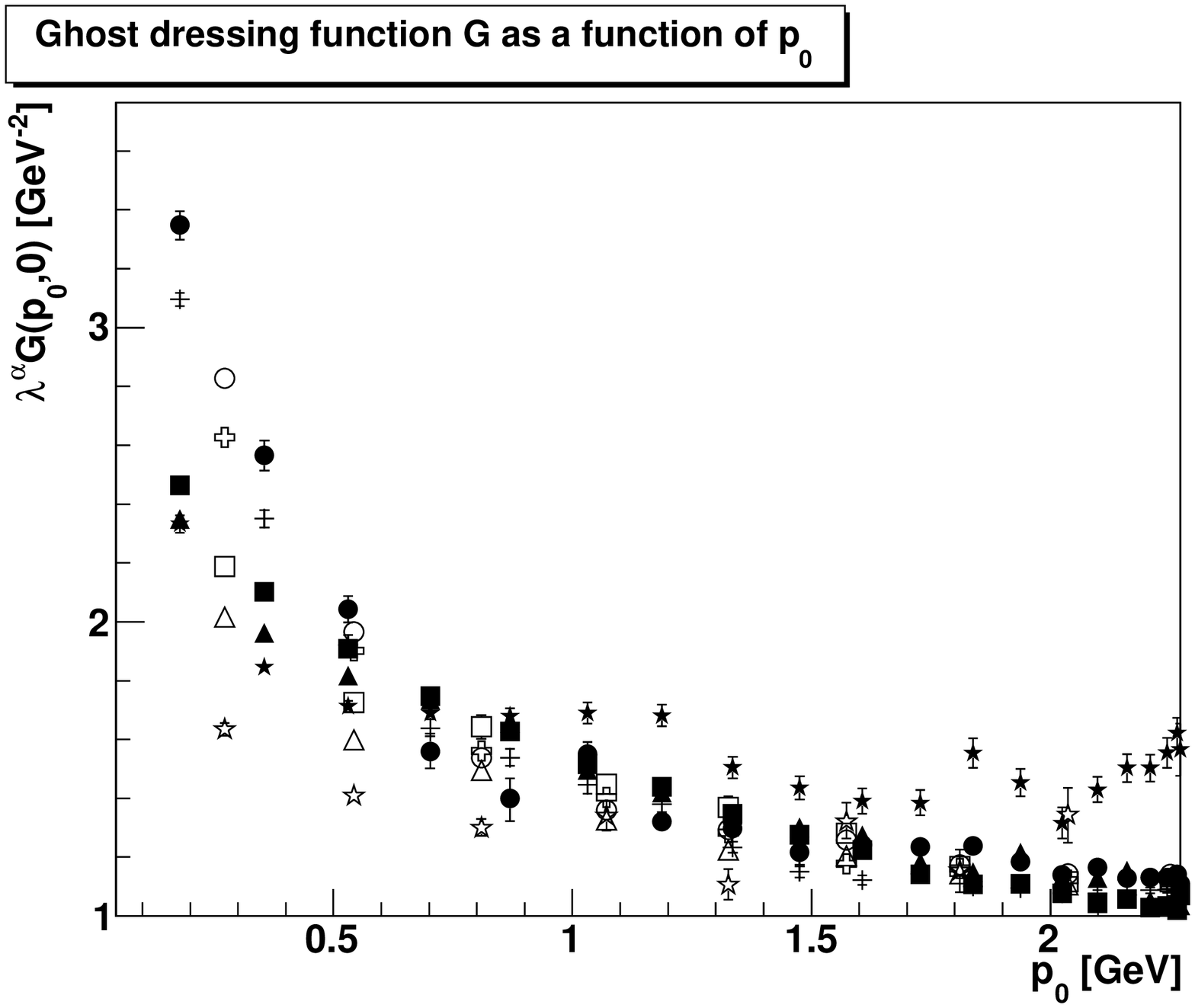}
\caption{\label{fghp-beta} The rescaled ghost dressing function $\lambda^\alpha G$ for different values of $\lambda$ for temporal momenta $p_0$, in the $3d$ case. Symbols used are the same as in the top row of Figure \ref{fghp}, with open symbols corresponding to $\beta=6.0$ and closed to $\beta=4.2$.}
\end{figure}

Finally, the {\bf ghost propagator} is shown in Figure \ref{fghp}. The dependence on $\lambda$ is clearly seen to increase with a decreasing ratio of $p_0/|\vec{p}|$. There is essentially no $\lambda$-dependence at $p_0=0$. Furthermore, as in Landau gauge, always a strong infrared enhancement is visible. Hence the most interesting case is the one of pure temporal momenta. In this case the propagator seems (to leading order) to be scaled by a constant factor, which is both $\lambda$- and $\beta$-dependent. This is tested explicitly in Figure \ref{fghp-beta}. Multiplying the dressing function by $\lambda^\alpha$, with $\alpha$ chosen to be 0.15 for $\beta=4.2$ and 0.1 for $\beta=6.0$, yields roughly agreeing dressing functions. This is a genuine  $\beta$-dependent effect, as the same exponents appear on a $20^3$ lattice at the same values of $\beta$. This may indicate that for this correlation function the scaling regime is moved as a function of $\beta$. The reason for this movement of the scaling region could be that the theory in Coulomb gauge is no longer finite, and renormalization has to be taken into account, even in three dimensions. If this should occur smoothly, the scaling region without renormalization should move to  larger $\beta$ with decreasing $\lambda$. An alternative explanation of this may be a combination of finite-size effects, leading seemingly to a $\beta$-dependence, and a genuine non-trivial dependence of the dressing function $G$ on $\lambda$. Such a non-trivial dependence is in fact found in functional calculations in the far infrared (see Appendix \ref{adse} and \cite{Fischer:2005qe}). However, if this is the correct interpretation, than this non-trivial $\lambda$-dependence seems to pertain to rather large momenta.

If the exponent $\alpha$ should be non-zero in the infinite-volume and continuum limit, then this would imply that the ghost propagator diverges for all $p_0$, independently of the spatial momentum, in the limit $\lambda\to 0$. This would be the behavior expected in the Gribov-Zwanziger scenario \cite{Gribov,zwanziger}.

It is not possible with the method used here to determine the ghost propagator in the case of $\lambda=0$, as the corresponding Faddeev-Popov operator is different from \pref{fpop}. In this case, it would be necessary to invert the Faddeev-Popov operator in the sub-space orthogonal not only to constant gauge transformations, as in Landau gauge \cite{Cucchieri:2006tf}, but also to all purely temporal gauge transformations. If, however, the limiting Coulomb gauge $\lambda\to 0$, the simple and the complete Coulomb gauges with $\lambda=0$ coincide also for the ghost propagator (as in the case of the gluon propagator), then an infrared-divergent, instantaneous ghost propagator would be obtained. This is the limiting behavior that is suggested from Figure \ref{fghp}.

\vskip 3mm
Concluding, the results show how the propagators evolve smoothly with $\lambda$ from a Landau-like behavior to a Coulomb-like behavior. However, below a certain momentum, which decreases with decreasing $\lambda$, the propagators always show qualitatively a Landau-like behavior. In particular, the temporal propagator as a function of spatial momenta shows a more and more diverging behavior, as in Coulomb gauge, at intermediate momenta, before becoming infrared-suppressed, as in Landau gauge. On the other hand, the transverse propagator as a function of transverse momenta becomes more and more infrared-suppressed, as it is in Coulomb gauge. Finally, the ghost propagator as a function of pure temporal momenta shows a rather pronounced dependence on $\lambda$ and $\beta$.

All in all, the propagators show a smooth development towards a Coulomb-like behavior in an intermediate momentum regime. Only in the far infrared does the original Landau-like behavior persist. The limit $\lambda\to 0$ thus corresponds to the limit in which the onset momentum of Landau-like behavior is moved towards zero momentum. As the Coulomb-like behavior and the Landau-like behavior are different, this limit is not smooth on the level of correlation functions. These results agree qualitatively with the naive momentum-rescaling argument of \cite{Cucchieri:2007uj}.

An interesting observation is the coincidence of the three different definitions of the Coulomb gauge. While the simple Coulomb gauge leaves the component $A_0$ completely unfixed, the limiting and the complete Coulomb gauge fix this component. However, in both cases the gluon propagators coincide within statistical errors. That is a consequence of the additional constraint not being a local condition on the fields. It only ensures $\pd_0 A_0=0$ on average, just as in the case of an unfixed temporal direction. To completely understand the properties of the limit of the interpolating gauge, this has to be investigated further, but is already in line with arguments from functional calculations \cite{Reinhardt:2008pr}.

\subsection{Propagators in 4 dimensions}\label{s4d}

The situation in four space-time dimensions turns out to be qualitatively the same as in three dimensions. Still, quantitatively, differences appear. One is due to the fact that finite-volume effects are harder to control, since calculations are in general more costly. The second is the need for renormalization even outside the Coulomb gauge: The theory is no longer finite in four dimensions. This point makes the situation a bit more intricate, as the gauge parameter $\lambda$ also gets renormalized by a renormalization constant $Z_\lambda$. This influences of course the definition of the ghost dressing function \pref{ghp}. In addition, the ghost dressing function itself gets renormalized by the wave-function renormalization constant $\widetilde{Z}_3$. The corresponding relative renormalization constants between two $\beta$-values will be determined by
\bea
\frac{\widetilde Z_3(\beta_1)}{\widetilde Z_3(\beta_2)}&=&\frac{D_G(0,|\vec p|,\beta_1)}{D_G(0,|\vec p|,\beta_2)}\label{zghost}\\
\frac{Z_\lambda(\beta_1)}{Z_\lambda(\beta_2)}&\approx&\frac{D_G(p_0,0,\beta_2)}{D_G(p_0,0,\beta_1)}\frac{D_G(0,|\vec p|,\beta_1)}{D_G(0,|\vec p|,\beta_2)}\label{lambdaren}.
\eea
\no The absolute normalization can be chosen at will, and is of no importance here. The second relation is based on \pref{ghp}, and would be exact only if the dressing function did not depend on $\beta$. Within the domain of perturbation theory, this dependence exists, but is only logarithmic. For the quite similar values of $\beta$ employed here, relation \pref{lambdaren} is hence an acceptable approximation at sufficiently large momenta.

\begin{table}[tbh!]
\caption{\label{trenorm}Ratios \prefr{zghost}{zgluon} of the renormalization constants. The matching condition was a continuous matching at the largest edge-momenta of the smaller $\beta$-value. Note that at $\lambda=1$ the gauge parameter is not renormalized because Landau gauge is a fixed-point of the renormalization group. At $\lambda=0$ no results on the ghost propagator are available, see text. At $\lambda=1$ the two gluon renormalization constants coincide trivially. Note that, at least in the simple Coulomb gauge, it is strictly speaking not possible to renormalize purely multiplicatively \cite{Watson:2008fb}.}
\begin{ruledtabular}
\vspace{1mm}
\begin{tabular}{c c c c c c }
$\lambda$ & $\beta_1$ & $\beta_2$ & $\frac{\tilde Z_3(\beta_1)}{\tilde Z_3(\beta_2)}$ & $\frac{Z_\lambda(\beta_1)}{Z_\lambda(\beta_2)}$ & $\frac{Z_3^\tr(\beta_1)}{Z_3^\tr(\beta_2)}\approx \frac{Z_3^0(\beta_1)}{Z_3^0(\beta_2)}$ \cr
\hline
1 & 2.2 & 2.5 & 1.12 & 1 & 1.23 \cr
%\hline
1 & 2.5 & 2.8 & 1.08 & 1 & 1.28  \cr
\hline
0.5 & 2.2 & 2.5 & 1.12 & 0.95 & 1.23 \cr
%\hline
0.5 & 2.5 & 2.8 & 1.09 & 0.95 & 1.28 \cr
\hline
0.1 & 2.2 & 2.5 & 1.12 & 0.65 & 1.23 \cr
%\hline
0.1 & 2.5 & 2.8 & 1.10 & 0.91 & 1.28 \cr
\hline
0.05 & 2.2 & 2.5 & 1.12 & 0.55 & 1.23 \cr
%\hline
0.05 & 2.5 & 2.8 & 1.11 & 0.83 & 1.28 \cr
\hline
0.01 & 2.2 & 2.5 & 1.12 & 0.4 & 1.23 \cr
%\hline
0.01 & 2.5 & 2.8 & 1.12 & 0.77 & 1.28 \cr
\hline
0 & 2.2 & 2.5 & & & 0.65 \cr
%\hline
0 & 2.5 & 2.8 & & & 0.74 \cr
\end{tabular}
\end{ruledtabular}
\end{table}

In case of the gluon propagator, the corresponding ratios of renormalization constants will be defined by
\bea
\frac{Z_3^\tr(\beta_1)}{Z_3^\tr(\beta_2)}&=&\frac{D^\tr(0,|\vec p|,\beta_1)}{D^\tr(0,|\vec p|,\beta_2)}\,,\label{zgluon}\\
\frac{Z_3^0(\beta_1)}{Z_3^0(\beta_2)}&=&\frac{D_{00}(0,|\vec p|,\beta_1)}{D_{00}(0,|\vec p|,\beta_2)}\,.\label{zgluon0}
\eea
\no Note that due to the lack of Euclidean invariance both propagators have to be renormalized separately \cite{Baulieu:1998kx,Burnel:2008zz}, and therefore independent wave-function renormalization constants $Z_3^\tr$ for the transverse gluon propagator and $Z_3^0$ for the temporal gluon propagator have to be introduced. This is in contrast to Landau gauge. Also, note that when reaching Coulomb gauge the combination $g^2D_{00}$ is a renormalization group invariant, and the corresponding inverse renormalization constant of the propagator alone yields the renormalization of the coupling constant \cite{Baulieu:1998kx}.

The approximate renormalization constants are given in Table \ref{trenorm}. The ratios for the transverse and the temporal part of the gluon propagator are found to be essentially equal, and thus only one is listed. Of course, these ratios cannot be determined precisely, as the same momenta are not available for different lattices. The renormalization has been performed to obtain a smooth connection at the largest edge-momentum of the lattice with the larger physical volume at the same lattice volume\footnote{The ultraviolet properties of the gluon propagator are only weakly affected by finite-volume effects. Hence the difference of physical volume can be ignored for the purpose of determining the renormalization constants at ultraviolet momenta.}. In addition, statistical fluctuations and systematic errors also prevent a precise measurement within the present scope, so these results are more qualitative than quantitative. However, the error on the ratio is of the same size as the statistical and systematic errors of the propagators, and thus a few, up to ten, percent for statistical errors and similarly for the systematic errors. The absolute normalization is chosen not to change the results at $\beta=2.5$.

These results indicate that the wave-function renormalization constants do not depend markedly on $\lambda$. However, the renormalization constant of $\lambda$ itself depends on $\lambda$. This was somewhat anticipated. When moving further away from a fixed point of the renormalization group, renormalization effects should become stronger and stronger with ``distance''. That Landau gauge is a fixed point is explicitly demonstrated by the decrease in $Z_\lambda$ with increasing $\beta$. From here on, only renormalized results will be shown.

Aside from these renormalization effects, the results in four dimensions show the same qualitative behavior as in three dimensions. However, this is only true when comparing roughly the same volumes, or, in fact, slightly smaller volumes in the three-dimensional case, to the volumes available in four dimensions. In particular, nearly none of the infrared effects visible in three dimensions can (yet?) be seen at the volumes used in four dimensions. One important exception below will be the transverse gluon propagator as a function of transverse momentum at very small $\lambda$. This has already been discussed previously in \cite{Cucchieri:2007uj}. Hence, here only a more compact presentation of the results will be given compared to the one in three dimensions in the previous section. In particular, as the results for non-vanishing energy and non-vanishing spatial momenta only interpolate smoothly between different behaviors if one of them is zero, here only these cases will be shown. Of course, in these kinematic configurations the results also look qualitatively similar to the three-dimensional case.

\begin{figure}
\includegraphics[width=\linewidth]{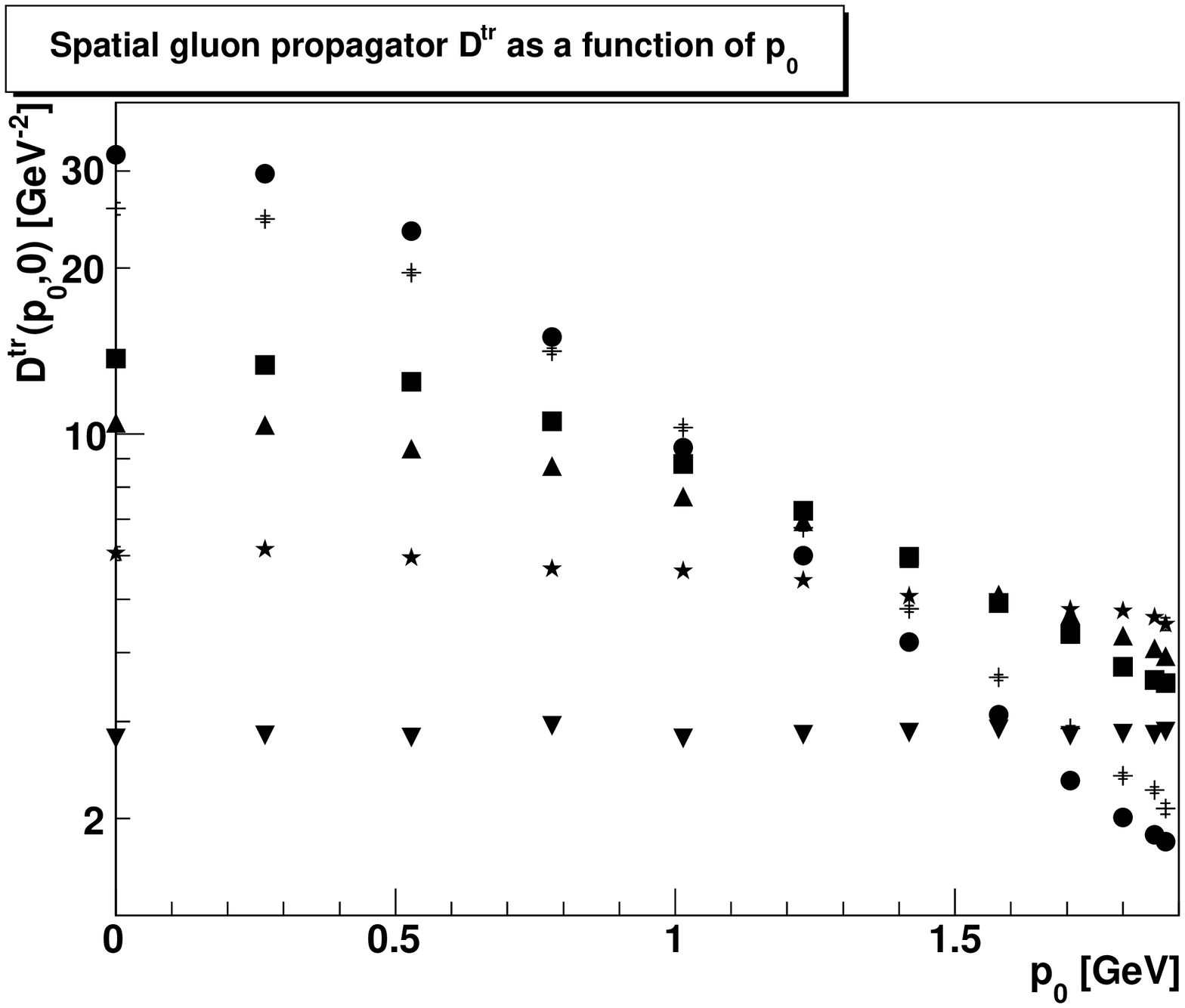}\\
\includegraphics[width=\linewidth]{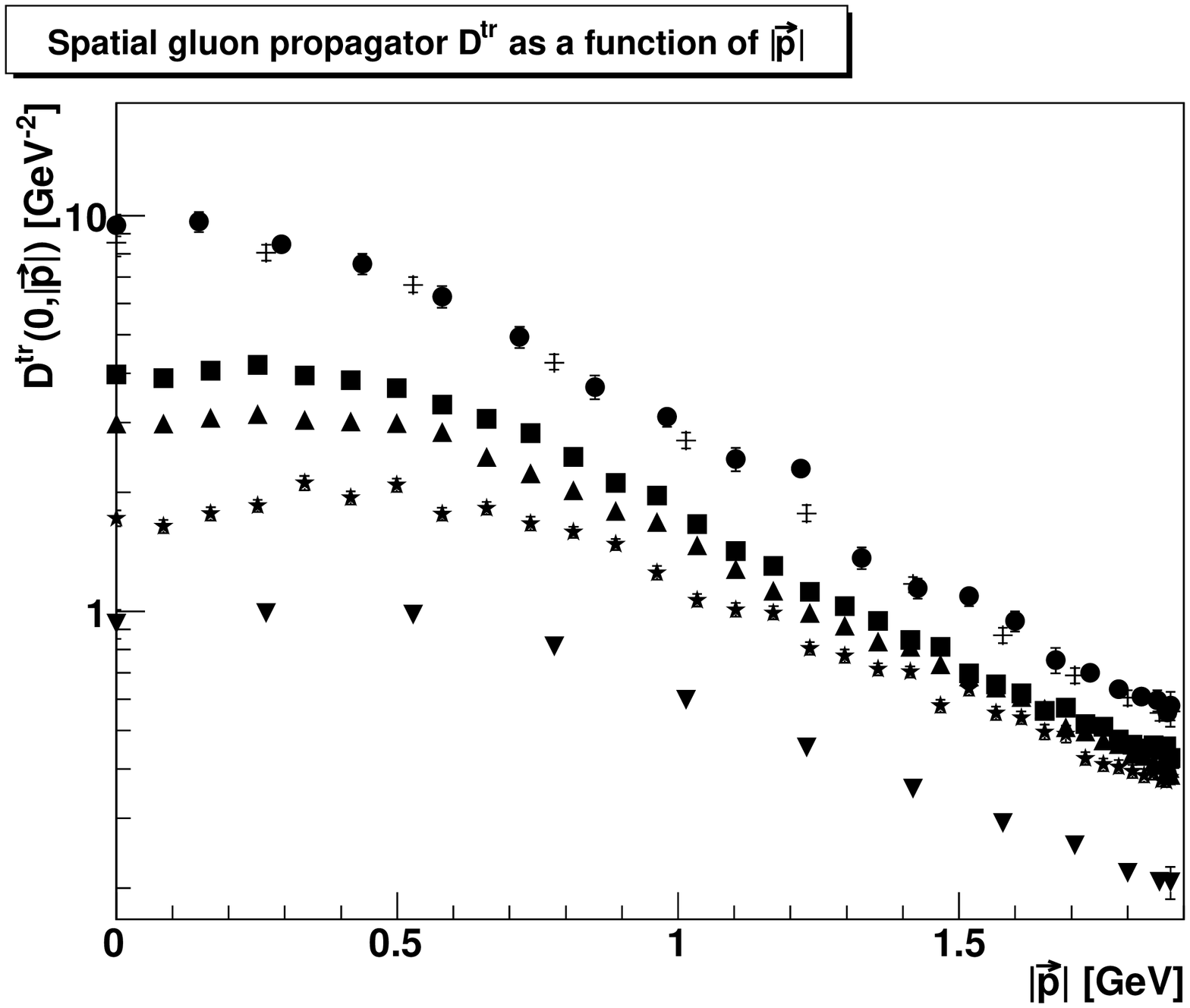}
\caption{\label{fdtr4}The transverse gluon propagator $D^\tr$ for different values of $\lambda$. The top panel is the propagator as a function of pure temporal momenta $p_0$, the bottom panel of pure spatial momenta $\vec p$. The spatial momenta are measured along the $x$-axis. Results shown are for $\lambda=1$, i.e.\ Landau gauge (circles), $\lambda=1/2$ (crosses), $\lambda=1/10$ (squares), $\lambda=1/20$ (triangles), $\lambda=1/100$ (stars), and $\lambda=0$ (upside-down triangles). All results are at $\beta=2.2$. Also, $\lambda=0$ corresponds to the simple-Coulomb-gauge case. Results at larger $\beta$ look similar, but are more strongly affected by finite-size effects. In case of the spatial momenta, the results for $\lambda=1$ come from a lattice of size $40^4$, while $\lambda=1/10$, $\lambda=1/20$, and $\lambda=1/100$ are from a lattice of size $70^4$. In all other cases, the lattice size was $22^4$.}
\end{figure}

\begin{figure}
\includegraphics[width=\linewidth]{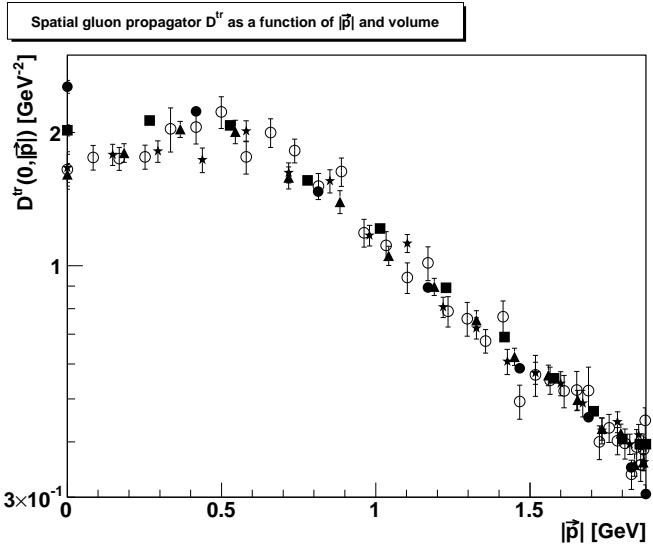}
\caption{\label{fdtr-v4} The transverse gluon propagator $D^\tr$ as a function of $|\vec{p}|$ at $\lambda=1/100$ for different volumes, in the $4d$ case. Circles, squares, triangles, stars, and open circles correspond to systems of size 14$^4$, 22$^4$, 32$^4$ (from \cite{Cucchieri:2007uj}), 40$^4$, and $70^4$, all at $\beta=2.2$.}
\end{figure}

The results for the transverse gluon propagator are shown in Figure \ref{fdtr4}. As a function of spatial
momentum $|\vec p|$, the propagator becomes more and more infrared-suppressed, as already observed for three
dimensions. No clear maximum is visible at large $\lambda$, owing to the small volumes in four dimensions. However, for the large volumes at values of $\lambda\le 0.1$, a shallow maximum is observed. This maximum moves to larger momenta with decreasing $\lambda$, as in the three-dimensional case. This can also be seen in a direct comparison of different volumes, shown in Figure \ref{fdtr-v4}. This confirms the results obtained in \cite{Cucchieri:2007uj}. The observed infrared-suppressed gluon propagator in Coulomb gauge is in accordance with the results of \cite{Burgio:2008jr,Cucchieri:2000gu,Nakagawa:2009zf}, though here the momentum-dependent aspects of renormalization in Coulomb gauge still have to be taken into account \cite{Watson:2008fb}.

The behavior as a function of temporal momentum $p_0$ is also reminiscent of the results in three dimensions.
But due to the restriction in volume the effects are even less pronounced. Still, it is visible how the
propagator becomes flatter with decreasing $\lambda$ and is constant at $\lambda=0$, except at zero momentum.

Hence, up to the effects of the smaller volumes and renormalization, the transverse gluon propagator behaves 
qualitatively as in three dimensions. In particular, as a function of spatial momentum, it becomes more
infrared-suppressed with decreasing $\lambda$.

\begin{figure}
\includegraphics[width=\linewidth]{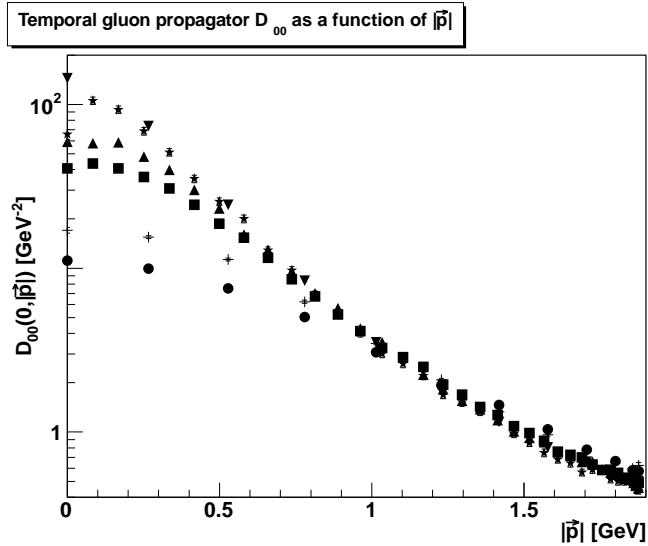}
\caption{\label{f4d-d00} The temporal gluon propagator $D_{00}$ for different values of $\lambda$ as a function
of pure spatial momenta $\vec p$, in the $4d$ case. The momentum is measured along the $x$-axis. Symbols have the same meaning as in Figure \ref{fdtr4}.}
\end{figure}

The results for the temporal gluon propagator are shown in Figure \ref{f4d-d00}. The situation is essentially the same
as in three dimensions. The propagator as a function of spatial momenta becomes more and more enhanced in the
infrared. Of course, for the volumes accessible here, it is not possible to check whether it then bends over
again in the far infrared, as it was observed in three dimensions in sufficiently large volumes. When adding a dependence also on the temporal momenta, 
it is found that the propagator increases with decreasing $\lambda$ as well. Note that this implies a much stronger falloff towards 
zero temporal momenta, which always occurs due to the gauge condition for vanishing spatial momenta.

The dependence on volume is less pronounced (and less spectacular so far) than in case of the spatial gluon propagator, 
and hence will not be shown explicitly.

\begin{figure}
\includegraphics[width=\linewidth]{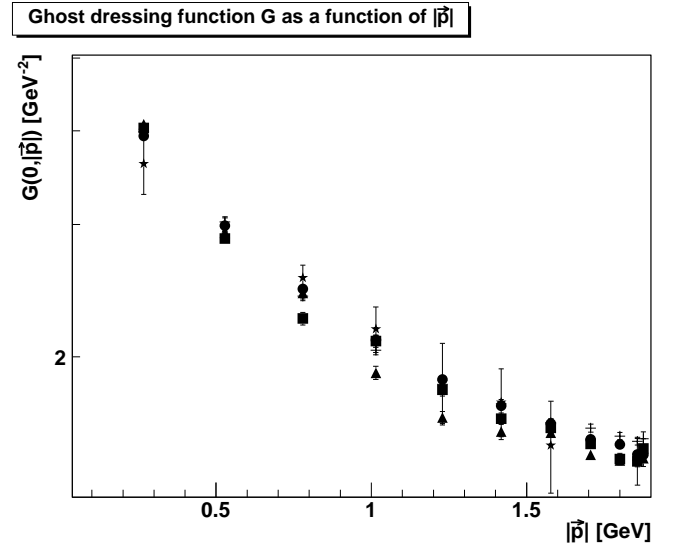}\\
\includegraphics[width=\linewidth]{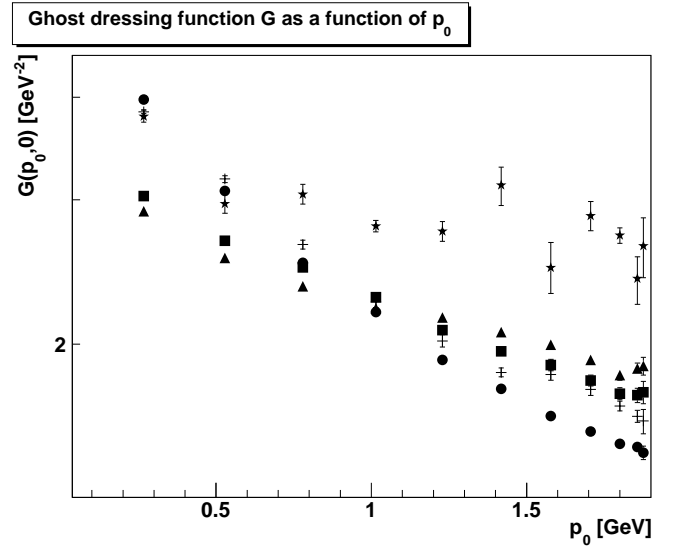}
\caption{\label{f4d-ghp} The ghost dressing function $G$ for different values of $\lambda$ as a function of pure spatial momenta $\vec p$, in the $4d$ case, in the top panel. The dependence on pure temporal momenta $p_0$ is given in the bottom panel. Results shown are for $\lambda=1$, i.e.\ Landau gauge (circles), $\lambda=1/2$ (crosses), $\lambda=1/10$ (squares), $\lambda=1/20$ (triangles), and $\lambda=1/100$ (stars), all at $\beta=2.2$ on a $22^4$ lattice. Results at larger $\beta$ are essentially identical.}
\end{figure}

Finally, the ghost dressing function is shown in Figure \ref{f4d-ghp}.
It shows a clearly infrared enhanced behavior at small momenta, essentially unaltered compared to Landau
gauge. Only in the pure temporal case is an additional increase with $\lambda$ for all momenta observed which is not
a pure renormalization effect, but instead similar to the situation in three dimensions.

As in three dimensions, it is possible to let all curves at $\vec p=\vec 0$ collapse by multiplying with $\lambda^\alpha$. In
contrast to three dimensions, the exponent $\alpha$ is essentially $\beta$-independent, as a consequence of renormalization. It increases slowly with $\lambda$, more 
or less leveling off at the smallest achieved $\lambda$ value at around $\alpha\approx0.08$. This rescaling leads to an apparent weaker 
divergence of the ghost dressing function. However, with such a small $\alpha$, the effect is rather weak.

Thus the results in four dimensions are, up to finite-volume effects, qualitatively the same as in three dimensions. When
decreasing $\lambda$ then at least at mid-momenta a Coulomb-like behavior is observed. In particular the transverse gluon
propagator as a function of spatial momentum becomes more infrared-suppressed, the temporal propagator
as function of spatial momentum becomes more mid-momentum enhanced. The ghost seems to be more or less inert,
but as it is the quantity usually least affected by changes which are effectively volume changes, this is not surprising.

\begin{figure}
\includegraphics[width=\linewidth]{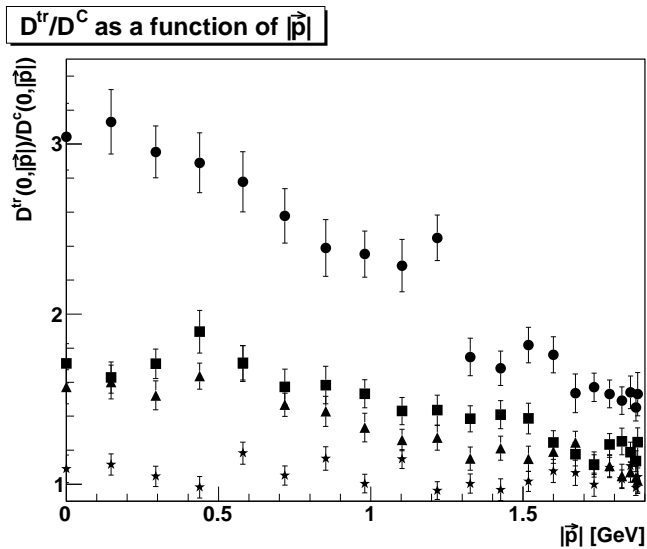}
\caption{\label{fratio}The ratio of the transverse gluon propagator \pref{dtr} to the instantaneous would-be Coulomb-gauge
propagator \pref{dinst} as a function of spatial momentum for $\lambda=1$ (circles), $\lambda=1/10$ (squares), $\lambda=1/20$
(triangles), and $\lambda=1/100$ (stars). The lattice was of size $40^4$ at $\beta=2.2$.}
\end{figure}

For Coulomb gauge the same applies, as was already said in the three-dimensional case: On the
level of the gluon propagators all definitions of the Coulomb gauge are indistinguishable. This approach can be made even quantitative, e.\ g.\ by comparing
the gluon propagator as a function of spatial momentum defined according to \pref{dtr} with the would-be Coulomb-gauge propagator, i.e.\ the time-average of \pref{dtr}
\be
D^{\mathrm{C}}(|\vec p|) \; = \; \sum_t 
     \left(\delta_{ij} - \frac{p_i p_j}{{\vec p\,}^2}\right)
     \frac{ < A_i^a(t,\vec p)  A_j^a(t,-\vec p) > }{(N_c^2-1) (d-2) V} \label{dinst} \; .
\ee
\no This comparison is made in Figure \ref{fratio}. It is explicitly visible how the transverse gluon propagator tends to the 
instantaneous one, in particular in the infrared. In fact, at the smallest non-vanishing momenta the rate of approach is roughly
proportional to the unrenormalized $\lambda$ beyond Landau gauge.

\section{Coulomb potential}\label{scopot}

One remarkable property of Coulomb gauge is the existence of an upper bound to the conventional Wilson potential in terms of a gauge-dependent quantity, the so-called color-Coulomb potential \cite{Zwanziger:2002sh}. This condition is often called informally the ``no-confinement-without-Coulomb-confinement'' one, and has been repeatedly studied \cite{Greensite:2004ke,Nakagawa:2006fk,Iritani:2011zg}.

The Coulomb potential can be straightforwardly defined by use of link variables as
\be
V_C(R,0)-E_0=-\left.\ln\left<\frac{1}{2}\tr\left(U_0(\vec{x},t) U_0^+(\vec{y},t)\right)\right>\right\vert_{R\equiv\mid \vec{x}-\vec{y}\mid}\label{cp},
\ee
\no where $E_0$ is a constant energy offset, which can be removed by renormalization of the potential. For convenience, it is here always chosen such that the Coulomb potential is zero at zero distance. Of course, this identification is only possible up to ${\cal{O}}(a^2)$, but studies at various $\beta$ indicated that the effects are small, but discernible. However, they affected essentially only the value of the string tension, but neither the functional shape of $V_C$, nor its dependence on $\lambda$, and an extensive analysis is therefore skipped.

\begin{figure}
\includegraphics[width=\linewidth]{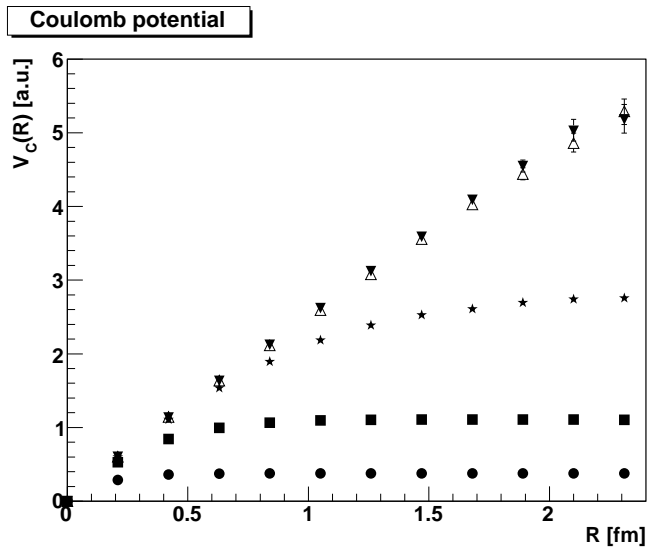}
\caption{\label{coulpot}The color-Coulomb potential \pref{cp} as a function of distance on a $24^4$ lattice at $\beta=2.2$. Other volumes and discretizations follow the expected pattern. The gauges are $\lambda=1$ (circles), $\lambda=1/10$ (squares), $\lambda=1/100$ (stars), simple Coulomb gauge (upside-down-triangles) and fully fixed Coulomb gauge (triangles).}
\end{figure}

The result is shown in Figure \ref{coulpot}. It is clearly visible that the color-Coulomb potential for all gauges, except Coulomb gauge, is asymptotically flat. However, with decreasing $\lambda$, a linear rise is seen initially at small distances. Finally, all three possible definitions of the Coulomb gauge, simple Coulomb gauge, fully-fixed Coulomb gauge, and the limit $\lambda\to 0$ coincide in case of the color-Coulomb potential.

This result corroborates therefore the finding that correlation functions in interpolating gauges at small distances behave Coulomb-gauge-like, and only at large distances revert back to a Landau-gauge behavior. Hence, at short range, the interpolating gauges are almost Coulomb-like. The notion of small is therefore necessarily set to be of order $1/(\lambda\Lambda_{\mathrm{YM}})$, though, of course, the functional shape could be different. This is only an estimate, but in line with the interpretation given in \cite{Cucchieri:2007uj}.

\begin{figure}
\includegraphics[width=\linewidth]{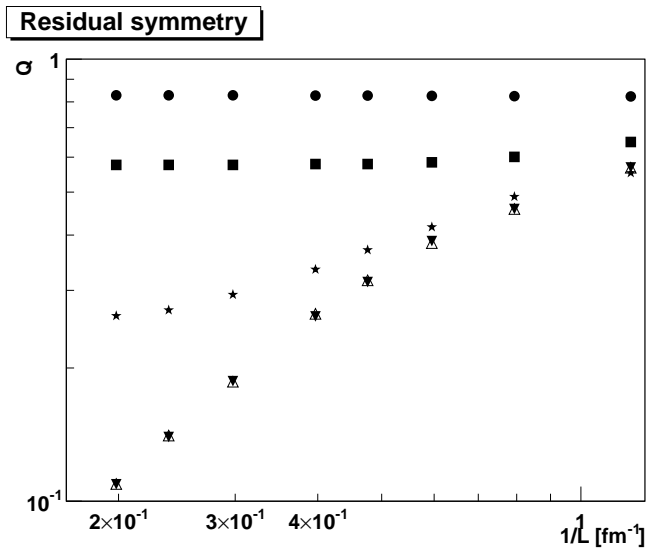}
\caption{\label{figq}The remnant-residual-symmetry order parameter $Q$ \pref{q} as a function of $L=aN_x$ at $\beta=2.2$. Though the pre-factors appearing are strongly $\beta$-dependent, as investigations at $\beta=2.5$ and $\beta=2.8$ have shown, the functional behavior remains the same. The gauges are $\lambda=1$ (circles), $\lambda=1/10$ (squares), $\lambda=1/100$ (stars), simple Coulomb gauge (upside-down-triangles) and fully fixed Coulomb gauge (triangles).}
\end{figure}

The fact that once more all three implementations of Coulomb gauge are the same is at this point not surprising. However, there exists an order parameter $Q$ for the residual gauge freedom of Coulomb gauge, defined as \cite{Greensite:2004ke}
\bea
Q&=&\frac{1}{N_t}\sum_{n=1}^{N_t}\left<\sqrt{\frac{1}{2}\tr\tilde{U}(t)\tilde{U}(t)}\right>\label{q}\\
\tilde{U}(t)&=&\frac{1}{N_xN_yN_z}\sum_{x,y,z}U_0(x,t)\nn.
\eea
\no If for $N_xN_yN_zN_t\to\infty$ $Q$ is vanishing, the residual gauge symmetry is unbroken. Since it has been shown that this is equivalent to a non-zero Coulomb string-tension \cite{Greensite:2004ke}, it is possible to anticipate from Figure \ref{coulpot} already the result that $Q$ will vanish for all definitions of the Coulomb gauge. In fact, this is seen in the results presented in Figure \ref{figq}. In all definitions of the Coulomb gauge $Q$ vanishes with volume. This also implies that although the residual gauge symmetry has been fixed in the fully-fixed version of Coulomb gauge and by taking the limit of the interpolating gauge, the residual gauge symmetry is not broken on average. On the other hand, for all gauges except Coulomb gauge it is once more visible how for decreasing $\lambda$ at small volumes first a Coulomb-like pattern emerges before the original Landau-gauge pattern, with broken residual gauge symmetry, becomes manifest.

It should be noted that\footnote{We are grateful to Daniel Zwanziger for pointing this out.} $Q=0$ implies $\tilde{U}(t)=0$, i.~e., a zero matrix. The second type of Coulomb gauge, in which the temporal gauge freedom of Coulomb gauge is fixed in a Landau-like manner, is actually the condition to minimize \cite{Cucchieri:2007uj}
\be
\sum_{x,y,z,t}\tr\;U_0(x,t)\propto\sum_t \tr\;\tilde{U}(t)\nn.
\ee
\no In the confining case, this sum is thus always exactly zero in the infinite-volume limit, and therefore the corresponding gauge-fixing step is trivial. Thus, in the infinite-volume limit the fully-fixed Coulomb gauge and the simple Coulomb gauge coincide. That this appears to be also the case at finite volume, as found here, appears non-trivial.

\section{Summary}\label{ssum}

Summarizing, a systematic investigation of correlation functions of Yang-Mills theory in the class of $\lambda$-gauges interpolating between the Coulomb and the Landau gauge in three and four dimensions has been performed.

The main result is threefold. On the one hand, the gauge parameter interpolating between both gauges parametrizes a certain length scale. For distances shorter than this scale, all the determined correlation functions show an essentially Coulomb-like behavior, at least up to momenta where perturbation theory becomes the dominant contribution. For distances longer than this scale, the behavior is essentially that as in the Landau gauge. There is no indication that this separation of Coulomb-like and Landau-like behavior will change for larger volumes and/or finer discretizations, even though the asymptotic infrared behavior in Landau gauge is still sensitive to such changes.

Furthermore, the investigation of different realizations of the Coulomb gauge, with and without fixing the residual gauge degree of freedom, did not show any effect for any of the correlation functions. This is in accordance with the reasoning of \cite{Cucchieri:2007uj} and of the continuum investigations in \cite{Reinhardt:2008pr}.

Finally, the evolution of correlation functions with this gauge parameter from Landau gauge to Coulomb gauge is smooth, with the only possible exception of the point at zero four-momentum for the temporal gluon propagator: In the interpolating gauge it is forced to vanish there, while in Coulomb gauge it is expected to diverge. However, for any non-zero four-momenta, its behavior is smooth.

Thus the interpolating gauge indeed turns into an interpolation of scales. The results interpolate in the far infrared from a Landau-like behavior over a Coulomb-like behavior to the perturbative $\lambda$-like behavior. Hence the name of interpolating gauges is truly justified.

Concluding, it is remarkable that all energy-dependence is actually vanishing in all definitions of the Coulomb gauge. Thus, all investigated correlation functions become time-independent, and therefore do not describe propagating degrees of freedom, in contrast to physical (bound) states. In this sense, Coulomb gauge can be regarded as a physical gauge, for any of the investigated versions.

%%%%%%%%%%%%%%%%%%%%%%%%%%%%%%%%%%%%%%%%%%%%%%%%%%%%%%%%%%%%%%%%%%%%%%%%%%%%%%%%%%%%%%%%%%%%%%%%%%%

\acknowledgments

We are grateful to Attilio Cucchieri for helpful discussions and to Daniel Zwanziger for remarks on the manuscript. A.\ M. was supported by the DFG under grant numbers MA 3935/1-1, MA 3935/1-2, and MA 3935/5-1, and by the FWF under grant numbers P20330 and M1099-N16. T.\ M. was partially supported by FAPESP and by CNPq. \v{S}.\ O. was supported by the Slovak Grant Agency for Science, project VEGA No.\ 2/0070/09, by ERDF OP R$\&$D, project CE meta-QUTE ITMS 26240120022, and via Center of Excellence SAS QUTE. The ROOT framework \cite{Brun:1997pa} has been used in this project. Part of our simulations were done on the IBM supercomputer at S\~ao Paulo University (FAPESP grant 04/08928-3).

%%%%%%%%%%%%%%%%%%%%%%%%%%%%%%%%%%%%%%%%%%%%%%%%%%%%%%%%%%%%%%%%%%%%%%%%%%%%%%%%%%%%%%%%%%%%%%%%%%%

\appendix

\section{Infrared analysis using Dyson-Schwinger equations}\label{adse}

In this appendix, connection to the commonly used framework for interpolating gauges in functional methods will be made. Furthermore, it will be shown that the persistence of the Landau-like behavior in the far infrared (at least for non-perturbative $\lambda$-gauges) allows for a scaling solution not only in four dimensions, as shown in \cite{Fischer:2005qe}, but also in three dimensions.

In functional calculations, it is convenient to use instead of the quantities $D_{00}$ and $D^\tr$, defined in equations \prefr{d00}{dtr} above, different ones. Writing the gluon propagator as (no summation implied)
\bea
D_\mn&=&\left(\delta_{\mu 0}\delta_{\nu 0}-\lambda\delta_{\mu 0}(\delta_{\nu 0}-1)\frac{p_0 p_\nu}{\vec p^2}\right.\nn\\
&-&\lambda(\delta_{\mu 0}-1)\delta_{\nu 0}\frac{p_0 p_\mu}{\vec p^2}\\
&+&\left.\lambda^2(\delta_{\mu 0}-1)(\delta_{\nu 0}-1)\frac{p_\mu p_\nu p_0^2}{\vec p^4}\right)d_{00}\nn\\[2mm]
&&+(\delta_{\mu 0}-1)(\delta_{\nu 0}-1)\left(\delta_\mn-\frac{p_\mu p_\nu}{\vec p^2}\right)d^\tr,\nn
\eea
\no defines two scalar functions, $d_{00}$ and $d^\tr$.

Due to the appearance of $|\vec p|$ in the denominator, this possibility to parametrize the gluon propagator is not too useful in
lattice calculations. The scalar functions $d_{00}$ and $d^\tr$ are related to the ones defined in \prefr{d00}{dtr} by
\bea
d_{00}&=&D_{00}\nn\\
d^\tr&=&D^\tr-\lambda^2\frac{p_0^4}{\vec p^2(p_0^2+\vec p^2)}D_{00}\nn\\[2mm]
D^\tr&=&d^\tr+\lambda^2\frac{p_0^4}{\vec p^2(p_0^2+\vec p^2)}d_{00}.\nn
\eea
\no The main difference in the results between both definitions is that the maximum in spatial (temporal) directions\footnote{Spatial 
directions implies still a non-zero temporal component, as otherwise both definitions coincide.} is more pronounced
for $d^\tr$ ($D^\tr$) than for $D^\tr$ ($d^\tr$), when comparing both results with the lattice data. For the results presented in the main text, this is only a weak, quantitative effect.

The DSEs and their solutions can then be obtained along the same lines as in four dimensions \cite{Fischer:2005qe}. The equations are given by
\bea
D_G(p)^{-1}&=&\bar{p}p\label{idsegp}\\
&-&\frac{C_Ag^2}{(2\pi)^3}\int d^3k \bar{p}_\mu D_\mn(k) \bar{p}_\nu D_G(p+k)\nn\\
D_\mn(p)&=&D_{\mu\rho}(p)D^\tl_{\rho\sigma}(p)D_{\sigma\nu}(p)\nn\\
&+&\frac{C_Ag^2}{(2\pi)^3}\int d^3k D_{\mu\rho}(p)\bar{k}_\rho \bar{k}_\sigma D_{\sigma\nu}(p)\nn\\
&\times&D_G(k)D_G(p+k),\label{idseghp}\nn\\
\bar{p}_\mu&=&Hp_\mu=\left(p_0\lambda,\vec p\right)_\mu\nn,
\eea
\no using the metric $H=\diag(\lambda,1,1)$, and $\tl$ denotes a tree-level quantity. This is already an approximate system using the same truncation as in \cite{Fischer:2005qe}: Only tree-level terms and the self-consistent leading terms for a scaling solution, i.e., terms with at least one ghost-line, are retained. Furthermore, a perturbative color structure is assumed.

As in four dimensions \cite{Fischer:2005qe}, equations \prefr{idsegp}{idseghp} can be transformed so that they have the same form as in
Landau gauge. Since the gauge parameter is not renormalized in three dimensions, it is sufficient to introduce the quantities
\bea
\lambda d^3 p&=&d^3\bar{p}\nn\\
\bar{D}_\mn(\bar{p})&=&\frac{1}{\lambda} D_\mn(p)\nn\\
\bar{D}_G(\bar{p})&=&D_G(p)\nn.
\eea
\no Note that this implicitly defines in general rather complicated functions $\bar D_\mn$ and $\bar D_G$. Rewriting equations \prefr{dseghp}{dsegp} in terms of these new variables leads to
\bea
\bar{D}_G(\bar{p})^{-1}&=&\bar{p}p\label{dseghp}\\
&-&\frac{C_Ag^2}{(2\pi)^3}\int d^3\bar{k}\bar{p}_\mu\bar{D}_\mn(\bar{k})\bar{p}_\nu \bar{D}_G(\bar{p}+\bar{k})\nn\\
\bar{D}_\mn(\bar{p})&=&\bar{D}_{\mu\rho}(\bar{p})D^\tl_{\rho\sigma}(p){\bar D}_{\sigma\nu}(p)\label{dsegp}\\
&+&\frac{C_Ag^2}{(2\pi)^3}\int d^3\bar{k} {\bar D}_{\mu\rho}(\bar{p})\bar{k}_\rho \bar{k}_\sigma \bar{D}_{\sigma\nu}(\bar{p})\nn\\
&\times&\bar{D}_G(\bar{k})\bar{D}_G(\bar{p}+\bar{k}).\nn
\eea
\no Except for the tree-level terms, this system of equations looks now formally as in Landau gauge. In the gluon equation, \pref{dsegp}, the
tree-level term will turn out to be sub-leading, and can thus be neglected. In case of the ghost equation, this is in general more subtle.
However, when making an infrared ansatz of type
\bea
\bar{D}_G(\bar{p})&\sim& {\bar p}^{2\kappa-2}\nn\\
\left(\delta_\mn-\frac{\bar{p}_\mu\bar{p}_\nu}{\bar{p}^2}\right)\bar{D}_\mn(\bar{p})&\sim& {\bar p}^{2t-2}\nn,
\eea
\no the system needs to be renormalized to possess a solution. It is then possible to argue as in four dimensions \cite{Fischer:2005qe} that the tree-level term can be removed in the
renormalization process. Thus the system becomes completely equivalent to the Landau gauge case, and thus has the same two solutions \cite{Zwanziger:2003cf}
\bea
2\kappa+t&=&\frac{1}{2}\nn\\
\kappa&=&-\frac{1}{2}\nn\\
\kappa&=&-0.39\dots\nn\\
\eea
\no In particular, this leads to the infrared behavior for the original functions of
\bea
D_G(p)&\sim&\frac{1}{(\vec p^2+\lambda^2 p_0^2)^{1-\kappa}}\label{dseghpir}\\
d^\tr(p)&\sim&\frac{\lambda}{(\vec p^2+\lambda^2 p_0^2)^{1-t}}\nn\\
d_{00}(p)&\sim&\frac{\lambda^3\vec p^2}{(\vec p^2+\lambda^2 p_0^2)^{2-t	}}\nn.
\eea
\no The ghost diverges at zero $\bar{p}^2$. Its anomalous scaling with $\lambda$ for $\vec p=\vec 0$ is likely directly related to the appearance of the factor $\lambda^{2-2\kappa}$ in \pref{dseghpir}. The transverse gluon propagator is consistently suppressed from all directions. The temporal gluon propagator vanishes identically for $\vec p=\vec 0$. From all other directions it is also infrared-suppressed, but this is seemingly not so much due to the exponent $t$, but due to the explicit factor of $\vec p^2$ in the numerator. In fact, $2-t$ is not larger than zero.

%%%%%%%%%%%%%%%%%%%%%%%%%%%%%%%%%%%%%%%%%%%%%%%%%%%%%%%%%%%%%%%%%%%%%%%%%%%%%%%%%%%%%%%%%%%%%%%%%%%

\end{document}